\documentstyle[12pt]{article}
\newcommand{\ag}{\mbox{I \hspace{-0.82em} H}}
\def\bbox{{\,\lower0.9pt\vbox{\hrule \hbox{\vrule height 0.2 cm  

\hskip 0.2 cm 

\vrule  height 0.2 cm}\hrule}\,}}
\newcommand{\R}{\mbox{I \hspace{-0.82em} R}}
\def\bbox{{\,\lower0.9pt\vbox{\hrule \hbox{\vrule height 0.2 cm  

\hskip 0.2 cm 

\vrule  height 0.2 cm}\hrule}\,}}

\begin{document}

\setlength{\unitlength}{1mm}
\title{{\hfill {\small Alberta-Thy 03-98} } \vspace*{2cm} \\
Thermal Fields, Entropy, and Black Holes}
\author{\\
V.P. Frolov$^{1,2,3}$ and
D.V. Fursaev$^{1,4}$ \date{}}
\maketitle
\noindent  {
$^{1}${ \em
Theoretical Physics Institute, Department of Physics, \ University of
Alberta, \\ Edmonton, Canada T6G 2J1}
\\ $^{2}${\em CIAR Cosmology and Gravity Program}
\\ $^{3}${\em P.N.Lebedev Physics Institute,  Leninskii Prospect 53,
Moscow
117924, Russia}
\\ $^{4}${\em Joint Institute for
Nuclear Research, Laboratory of Theoretical Physics, \\
141 980 Dubna, Russia}
\\
\\
e-mails: frolov, dfursaev@phys.ualberta.ca
}
\bigskip

\begin{abstract}
In this review we describe  statistical mechanics of  quantum systems
in the presence of a Killing horizon and compare  
statistical-mechanical and one-loop contributions to black hole
entropy. Studying these questions was motivated by attempts to explain
the entropy of black holes as a statistical-mechanical entropy of
quantum fields  propagating near the black hole horizon.    We provide
an introduction to this field of research  and review its results.   In
particular, we discuss the relation between the statistical-mechanical
entropy of quantum fields and the Bekenstein-Hawking entropy in the
standard scheme with renormalization of gravitational coupling
constants and in the theories of induced gravity.
\end{abstract}

\bigskip

\baselineskip=.6cm

\newpage

\section{Introduction}
\setcounter{equation}0

Thermodynamics and statistical mechanics of black holes is one of the
most exciting and rapidly developing areas of black hole physics. 
Black holes are known to possess the properties similar to the
properties of thermodynamical systems \cite{BCH:73}.  According to this
analogy, a black hole has the  entropy $S^{BH}={1 \over 4G}{\cal A}$
where ${\cal A}$ is the surface area of its horizon and $G$ is the
Newton constant\footnote{We use units in which $\hbar=c=k_B=1$ ($k_B$
is the Bolzmann constant) and sign conventions of book \cite{MTW:73}
and, thus, we use the signature $(-,+,+,+)$ for the Lorentzian
metric.}. The quantity $S^{BH}$ was introduced in Refs.
\cite{Bekenstein:72}-\cite{Hawking:75} and is known as the {\it
Bekenstein-Hawking entropy}.  The temperature of a  black hole is
$T_H={\kappa \over 2\pi}$ where  $\kappa$ is the surface gravity of the
horizon. 

In the Einstein theory of general relativity the Bekenstein-Hawking
entropy is a pure geometrical quantity. The laws of black hole
thermodynamics are derived by making use of the classical Einstein
equations and rules of the differential geometry only.  If we compare
black holes with usual thermodynamical systems an important difference
can be easily observed: Black holes are nothing but an empty space with
a strong gravitational field while a usual body consists of material
constituents (atoms, molecules, etc.). Namely this  microscopical
structure  enables one to explain thermodynamical properties of the
body in terms of statistical mechanics of its constituents. Does a
black hole have  internal degrees of freedom which are  responsible for
the Bekenstein-Hawking entropy? This is a key question of black hole
physics.

The black hole entropy problem is important. The statistical-mechanical
derivation of the Bekenstein-Hawking entropy is a highly non-trivial
test for a fundamental theory of quantum gravity. Recent successful
calculations of $S^{BH}$  for extremal \cite{StVa:96}-\cite{MaSt:96}
and near extremal \cite{CaMa:96},\cite{HoSt:96} black holes in the
superstring theory clearly demonstrate this.  A review of these 
computations and further references can be found, for instance, in
\cite{Horo:97}, \cite{Peet:97}. Besides the superstring approach  there
are other approaches which attack the problem of black hole entropy
from  different directions. The explanation of the entropy of 3D black
holes  suggested by Carlip {\it et al.}
\cite{Carlip:95},\cite{Strominger:97} or consideration in the framework
of the loop quantum gravity \cite{ABCK:97} are some of them.  

The subject of this review is related to the  approach which suggests
to explain the black hole entropy in terms of its quantum excitations. 
This idea   was formulated first in
\cite{ZuTh:85},\cite{Hooft:85},\cite{TPM:86} and it has stimulated a
large number of publications. 

The properties of physical vacuum, especially in the presence of
gravity,  are nontrivial.  In the state of vacuum there always exist
zero-point fluctuations of physical fields. An observer at rest near a
horizon would register these zero-point fluctuations in the form of  a
thermal atmosphere  of a black hole \cite{TPM:86}-\cite{Laflamme:89}. 
Historically the first suggestions to relate the entropy of a black
hole to the entropy of its {\it thermal atmosphere} were made in works
by  Thorne and Zurek \cite{ZuTh:85} and by  t'Hooft \cite{Hooft:85}.
t'Hooft \cite{Hooft:85} estimated the thermal entropy by assuming  that
the red-shifted  temperature of the atmosphere is $T_H$ and showed that
the entropy is proportional to the horizon area ${\cal A}$. To avoid
the divergences t'Hooft assumed that fields vanish within some distance
near the  horizon (the corresponding model was called the  {\it ``brick
wall'' model}). If this distance is of the order of a Planck lengths
the  thermal entropy turns out to be comparable to $S^{BH}$.

The reason why the static observer near the black hole  sees the vacuum
as a mixed state is explained by the loss of the information about a
part of the  quantum system located inside a the black hole horizon.
Bombelli, Koul, Lee and Sorkin \cite{BKLS:86} and  Srednicki
\cite{Srednicki:93} have showed that even in a flat space when
observations in vacuum are restricted by a part of the system, the
entropy is not zero and is proportional to the surface area of the 
restricted region $\Omega$ (for a recent discussion see
\cite{MuSeKo:97}).   Similar result for the entropy was also
established for non-zero spin fields \cite{LaWi:95} and for the pure
sates  different from the vacuum \cite{BePi:96}. The  non-vanishing
entropy appears because   the ``observable'' and ``non-observable''
vacuum fluctuations are entangled (correlated) on the boundary of
$\Omega$. By taking into account these properties the authors  of 
Refs. \cite{BKLS:86},\cite{Srednicki:93} suggested to explain $S^{BH}$
as  the entropy of {\em entanglement} between quantum fluctuations 
propagating on the different sides of the horizon.  

Frolov and Novikov \cite{FrNo:93} proposed to relate the black hole
entropy to the degrees of freedom corresponding to quantum states in
the black hole interior.  The density matrix of these degrees of
freedom can be obtained by  averaging the quantum state of the complete
system over the states of fields located outside the black hole.  For
modes in the close vicinity to the horizon this density matrix is {\em
thermal}.  The particles are created in pairs, and only one of the
components can be created outside the horizon. A pair inside the black
hole is in a pure state and does not contribute to the entropy. For
this reason, the statistical mechanical entropy is connected with {\em
entanglement}, and it can be written in the form of summation over the
modes in the black hole {\em exterior}. In other words,  this approach
incorporates main features of the  earlier  approaches
\cite{ZuTh:85},\cite{Hooft:85},\cite{BKLS:86}. 

A remarkable property of a black hole is that its entaglement entropy,
and entropy connected with its thermal atmosphere coincide 
\cite{Bekenstein:94}-\cite{Jacobson:94b}. In what follows we refer to
this quantity as to {\em statistical-mechanical entropy}.

Small fluctuations of fields (including the gravitational one)
propagating in the black hole background can be related to small
deformations of the black hole geometry. This can be explicitly
demonstrated in the  approach using the no-boundary wavefunction of the
black hole \cite{BFZ:95}. For this reason, counting states of quantum
fields is connected with the counting the states of quantum excitations
of the black hole.

In the general case, the relation of the statistical-mechanical entropy
and the "observable" thermodynamical entropy of the black hole is
highly nontrivial  \cite{Frolov:95}. The quantum fields near a black
hole have an important property. Namely, the density of levels of
quantum  mechanical Hamiltonians of the fields blows out near the
horizon. This results in the  divergence of the statistical-mechanical 
entropy. Susskind and Uglum \cite{SuUg:94} and Callan and Wilczek
\cite{CaWi:94} suggested   that this divergence   is related to the
ultraviolet one-loop divergences of the theory and it can be removed by
renormalizing  the Newtonian coupling constant.  This observation was
strongly supported by the result of Demers, Lafrance and  Myers
\cite{DLM:95} who showed that  the standard Pauli-Villars method 
regularizes the divergences of the statistical-mechanical  entropy on
the Reissner-Nordstrom background. Moreover in this regularization, the
entropy divergence for a minimally coupled scalar field  is  completely
eliminated by the standard ultraviolet  renormalization. The same
result for general  static black-hole backgrounds was proven by Fursaev
and Solodukhin \cite{FS:96}. It was done  by using the Euclidean
formulation of the theory with  conical singularities. 

Although the observation made in  \cite{SuUg:94}, \cite{CaWi:94} is
very important, there are two problems. First, in order to carry out
the renormalization one must introduce an infinite (negative) bare
entropy, which has no statistical-mechanical origin. Second,  in the
presence of non-minimal coupling of a field with the curvature, the
divergence of the statistical-mechanical  entropy and the ultraviolet
divergence are different \cite{DLM:95}--\cite{HKN:97}. 

Jacobson \cite{Jacobson:94} pointed out that  the first problem can be
solved if the Einstein gravity itself appears as a result of quantum
effects. A suitable simple example is  Sakharov's theory of induced
gravity \cite{Sakh:68},\cite{Sakh:76}. The models of induced gravity
which enable one to  check this hypothesis were constructed by Frolov,
Fursaev and Zelnikov  \cite{FFZ:97}. It was demonstrated that the
presence of the non-minimal coupling is necessary in order to get a
finite induced Newton constant. Moreover,  it makes it possible a
consistent derivation of the black hole degeneracy by counting the
degrees of freedom of constituents  \cite{FF:97a},\cite{FF:97b}.

\bigskip

The idea to relate $S^{BH}$ to quantum excitations of a black hole and
the pioneering papers \cite{ZuTh:85}, \cite{Hooft:85}, \cite{BKLS:86}, 
\cite{Srednicki:93}, \cite{FrNo:93} stimulated  the study of
statistical mechanics of quantum fields  in the  presence of a Killing
horizon.  More than  hundred papers was written on this subject. Our
review of these works has two  main aims: \begin{enumerate} \item To 
describe the statistical mechanics of quantum fields  in the  presence
of a Killing horizon, its methods and obtained results; \item To 
discuss the relation between the statistical mechanical entropy and the
observable black hole entropy $S^{BH}$. \end{enumerate} 

As an application of these results, we  present the
statistical-mechanical derivation of the black hole entropy in the
models of induced gravity. 

\bigskip

We begin with two formulations of finite-temperature  quantum theory on
static spacetimes which have been discussed in the literature.

The first is the {\it canonical} formulation  based on the
$(3+1)$-split of the spacetime. It allows one to define the free energy
$F^C$ of the system in terms of the one particle spectrum. The
advantage of the canonical formulation is that statistical-mechanical 
entropy $S^C$ can be derived from $F^C$ by the standard rules.

The second is the {\it covariant Euclidean} formulation. It starts with
the calculation of the one-loop effective action $W$ on the
Gibbons-Hawking instanton. The free energy $F^E$ in this approach is
$F^E=\beta^{-1}W$, where $\beta$ is the inverse temperature.   The
covariant Euclidean  formulation\footnote{The terms "canonical" and 
"covariant Euclidean" reflect the form of the presentation  of the
corresponding partition functions (in form of summing over the energy
levels and as a functional  integral, respectively). It should be
emphasized that there is no standard terminology.  We  use superscripts
$C$ and $E$ to refer to the quantities calculated in the canonical and
covariant Euclidean formulations.} is especially useful in application
to black hole thermodynamics. 

These two formulations are logically different, and in the presence of
a black hole their comparison  is non-trivial. The first part of the
review is devoted to this problem.  In Section 2 we define $F^E$ and
$F^C$ on spacetimes without horizons  and show the equivalence of these
functionals. In Section 3 we discuss the features of quantum  systems
related to the horizon.  Section 4 reviews the results of the canonical
formulation for the spacetimes with horizons. Special attention is
payed to divergences connected with the presence of the horizon and
methods of their regularization.  The covariant Euclidean formulation
in the presence of a horizon is given in Section 5.  We discuss
ultraviolet one-loop divergences and show that the divergent parts of
$F^E$ and $F^C$ are identical in the covariant regularizations, such as
Pauli-Villars and dimensional ones.

In the second part of the review we discuss the relation between 
statistical-mechanical entropy   and the observable thermodynamical
entropy of a black hole.  In Sections 6,7 we  demonstrate how the
divergencies of statistical-mechanical entropy are removed by the
renormalization of gravitational couplings in the tree-level black hole
entropy.  The presence of the bare classical entropy  make it
impossible to give a statistical-mechanical explanation of  the black
hole entropy in the theories which require the ultraviolet 
renormalization. This difficulty does not exist in the theories of
induced gravity. The mechanism of generation of the Bekenstein-Hawking
entropy in such theories is described in Section 8.   Our summary and
conclusions are represented in Section 9.  The Appendix  is devoted to
interpretation of the Noether charge which appears in the
renormalization formula for the entropy because of nonminimal 
couplings of the fields.

\section{Statistical mechanics of quantum fields 
in a static gravitational field without horizons}

\subsection{Description of the system}
\setcounter{equation}0

We begin with discussion of statistical  mechanics of quantum fields 
in a static gravitational field without horizons. This theory has been
formulated and investigated in  a number of papers starting with
pioneering works by  Gibbons \cite{Gibbons:77},\cite{Gibbons:78},
Gibbons and Perry \cite{GiPe:78},   Dowker and Kennedy \cite{DoKe:78}
in the early seventies. A typical example is a quantum field at finite
temperature in the gravitational field of a static non-rotating star.
The gravitational field is described by the metric
\begin{equation}\label{2.13}
ds^2=g_{00}dt^2 + g_{ab}dx^a dx^b~~~,
~~~a,b=1,2,3~~~,
\end{equation}
where $g_{00}(x)<0$. The metric (\ref{2.13}) depends only on the
spatial coordinates $x^a$, and so it admits the Killing field 
$\zeta=\partial /\partial t$.  On the spatial infinity the background
is asymptotically flat and, by assumption, a time component of the
metric $g_{00}$ tends to $-1$. Because metric (\ref{2.13}) does not
depend on time, one can easily define  the statistical-mechanical
ensembles of different fields on this background. We will be dealing
with canonical ensembles  characterized by the temperature
$T=\beta^{-1}$ measured at asymptotic infinity\footnote{ It is well
known that an infinite bath of thermal radiation is a gravitationally
unstable system. In order to deal with a stable situation one may
consider a cavity of a finite size filled with the radiation. To
characterize the system one can fix the temperature at the boundary or
use the red-shifted  temperature at infinity. These remarks are also
valid  for the case of a black hole being in thermal equilibrium with
the radiation. For a non-rotating black hole of mass $M$ the size of
the cavity must be  less than $3M$ \cite{York:86}. Speaking about the
canonical ensemble we always assume that such a boundary exist.  Our
main subject is properties of ensembles of quantum fields connected
with the presence of the horizon.  The presence of external boundaries
is not important for this consideration. For this reason we are not
discussing them in more details.}.  The local Tolman  temperature
measured by an observer at a point $x^a$ is
$T_{loc}=|g_{00}|^{-1/2}\beta^{-1}$.

To investigate the cases of  Bose and Fermi statistics we will be
considering, as an example, free scalar  ($\phi$) and Dirac ($\psi$)
fields. The fields obey the  Klein-Gordon and Dirac equations
\begin{equation}\label{2.18}
(-\nabla^\mu\nabla_\mu+m^2+\xi R)\phi=0~~~,
\end{equation}
\begin{equation}\label{2.19}
(\gamma^\mu\nabla_\mu+m)\psi=0~~~,
\end{equation}
respectively. Here $R$ is the scalar curvature and $\xi$ is the
parameter of the non-minimal coupling.  The covariant derivatives
$\nabla_\mu$ are defined  according with the spin of the field.  The
Dirac $\gamma$-matrices  $\gamma^\mu=(\gamma^0,\gamma^\alpha)$ are
defined  by the standard relations
$\{\gamma^\mu,\gamma^\nu\}=2g^{\mu\nu}$. Note  that $\gamma^0$ is
anti-Hermitean matrix. We define the spinor  derivative as
$\nabla_\mu=\partial_\mu+\Gamma_\mu$, where  $\Gamma_\mu= \frac
18[\gamma^\lambda,\gamma^\rho]~V^l_\rho \nabla_\mu V_{l\lambda}$ is the
connection  and $V^l_\nu$ are the tetrads.

The following scalar products defined for solutions of equations
(\ref{2.18}) and (\ref{2.19})
\begin{equation}\label{2.18a}
< \phi_1,\phi_2>=i\int_{\cal B}~d^3x~\sqrt{^{(3)}g|g_{00}|^{-1}}
\left(\phi_{1}^{*}\,\partial_t\phi_{2}-
\phi_{2}\,\partial_t\phi_{1}^{*}\right)
~~~,
\end{equation}
\begin{equation}\label{2.19a}
< \psi_{1},\psi_{2}>=\int_{\cal B}~d^3x~\sqrt{^{(3)}g}\, 
\psi_{1}^{+}\psi_{2}~~~,
\end{equation}
where $^{(3)}g=\det g_{ab}$, are independent of the choice 
of the total Cauchy surface ${\cal B}$.

\subsection{Canonical formulation and 
single-particle spectrum}

A canonical ensemble at temperature $\beta^{-1}$ is determined by the
partition function  \begin{equation}\label{2.24}
Z^C(\beta)=\mbox{Tr}~e^{-\beta:\hat{\cal{H}}:}~~~. \end{equation} The
operator $:\hat{\cal{H}}:$ is the  Hamiltonian of the secondary
quantized field.  It determines a unitary evolution of the quantum
field with respect to the Killing time $t$.  As usual we use the
decomposition of field operators onto positive and negative
frequencies   with respect to the time $t$ in order to define the 
creation and annihilation operators. The normal ordering in
(\ref{2.24}) is with respect to these operators.  For this normal
ordering  the energy of the state with zero temperature (vacuum)
vanishes. If the vacuum energy is non-trivial, definition (\ref{2.24})
can be easily modified. We will discuss this modification later. The
{\it canonical} free energy of the system is 
\begin{equation}\label{2.25}
F^C(\beta)=-\beta^{-1}\ln Z^C(\beta)~~~.
\end{equation}

To proceed with the computation  of $F^C(\beta)$ it is convenient  to
rewrite Eq. (\ref{2.25}) in another equivalent form based on a
single-particle spectrum. Let $\omega$ be  frequency of a field mode
with respect to the Killing time $t$. We  call the set of these
frequencies the  {\it single-particle spectrum}. The spectrum of
$\omega$ is uniquely defined by the boundary conditions  imposed on the
system. If the system is given in a  finite region with Dirichlet or
other conditions on the boundary the spectrum is discrete. Some
frequencies in this case can coincide and   we introduce the
corresponding degeneracy factor $d(\omega)$. Then Eq. (\ref{2.25}) can
be identically rewritten in the form \cite{LaLi}, \cite{Allen:86}
\begin{equation}\label{2.1a}
F^C[\beta]=
\eta\beta^{-1}\sum_\omega d(\omega)\ln{(1-\eta e^{-\beta\omega})}~~~.
\end{equation}
The factor $\eta$ is related to the statistics, and it takes values
$\eta=1$ and $\eta=-1$ for  Bose and Fermi fields, respectively. 
Equation (\ref{2.1a}) is well-defined: the degeneracies $d(\omega)$ of
the three-dimensional elliptic operators grow as $\omega^2$ at large
$\omega$  but due to the exponential cutoff series (\ref{2.1a})
converges. 

When the system has the infinite size the spectrum  of $\omega$ is
fixed by the appropriate asymptotic conditions. A usual requirement is
that fields fall fast enough at the spatial infinity. In this case the
spectrum is continuous and  the sum in (\ref{2.1a}) has to be replaced
by the integral
\begin{equation}\label{2.1ab}
F^C[\beta]=
\eta \beta^{-1}\int_0^{\infty}d\omega {dn (\omega)
\over d\omega}\ln{(1-\eta e^{-\beta\omega})}~~~.
\end{equation}
The quantity ${dn (\omega)
\over d\omega}d\omega$ is the number of 
levels in the interval $(\omega,\omega+d\omega)$. 
Equation (\ref{2.1ab}) can be obtained from (\ref{2.1a})
in the limit when the interval between the levels 
shrinks to zero.

\subsection{Single-particle Hamiltonian}

The single-particle spectrum can be found from the
wave equations of the fields, 
Eqs. (\ref{2.18}), (\ref{2.19}).
On the static space (\ref{2.13})
these equations can be rewritten in the ``$3+1$''
form
\begin{equation}\label{2.2}
(\partial_t^2+H_s^2)\phi=0~~~,
~~~
(i\partial_t-H_d)\psi=0~~~,
\end{equation}
where the subscripts $s$ and $d$ are used for
scalar fields and Dirac spinors, respectively.
$H^2_s$ and $H_d$ are three-dimensional differential
operators
\begin{equation}\label{2.3}
H_s^2=
|g_{00}|(-\nabla^a\nabla_a -
w^a\nabla_a+m^2+\xi R)~~~,
\end{equation}
 \begin{equation}\label{2.4}
H_d=-i\gamma_0\left[\gamma^a
(\nabla_a+\frac 12 w_a)+m\right]~~~.
\end{equation}
The 3-dimensional covariant derivative $\nabla_a$  is defined in terms
of the metric $g_{ab}$  on the hypersurface of constant time
$t=\mbox{const}$.  We denote the constant time surface as ${\cal B}$.
Operations with the index $a$ are performed with the help of  3D-metric
$g_{ab}$ on ${\cal B}$, see Eq. (\ref{2.13}), and $w_a=\frac 12 
\nabla_\alpha\ln |g_{00}|$ is the three-dimensional part of the
acceleration vector $w_\mu=(0,w_a)$ of the Killing observer.

The operators $H_s$ and $H_d$  are {\it quantum-mechanical
single-particle Hamiltonians} because their eigen-values determine
single particle spectra. By substituting   the wave functions  with the
fixed energies 
$\phi(t,{\bf{x}})=e^{-i\omega t}\phi_{\omega}({\bf{x}})$,
$\psi(t,{\bf{x}})=e^{-i\omega t}\psi_{\omega}({\bf{x}})$
into Eqs. (\ref{2.2})
we arrive at the eigen-value problems
\begin{equation}\label{2.2'}
H_s^2\phi_\omega({\bf{x}})=\omega^2\phi_\omega({\bf{x}})~~~,~~~
H_d^2\psi_\omega({\bf{x}})=\omega^2\psi_\omega({\bf{x}})~~~.
\end{equation} 
It is easy to check that $H_s$ and $H_d$ are  
Hermitean operators
with respect to the following inner products 
 \begin{equation}\label{product2}
(\phi_{1},\phi_{2})=\int_{\cal B}~d^3x~\sqrt{^{(3)}g|g_{00}|^{-1}}
\,\phi_{1}^{*}({\bf{x}})\,\phi_{2}({\bf{x}})~~~,
\end{equation}
\begin{equation}\label{product}
(\psi_{1},\psi_{2})=\int_{\cal B}~d^3x~\sqrt{^{(3)}g} 
\,\psi_{1}^{+}({\bf{x}})\,\psi_{2}({\bf{x}})~~~,
\end{equation}
where 
$\psi^{+}$ denotes a Hermitean conjugated
spinor. 
The form of relations  (\ref{product2}) and (\ref{product})
follows from  inner products
(\ref{2.18a})--(\ref{2.19a})
for four-dimensional fields.
The inner products (\ref{product2}) and (\ref{product}) are used to
normalize the modes.

For the convenience of the computations we will 
also use another
{\it representation} of $H_s^2$ and $H_d$ 
which can be obtained by the following
transformation of functions and operators
\begin{equation}\label{2.32}
\bar{\phi}=e^{-\sigma}\phi~~~,~~~
\bar{\psi}=e^{-\frac 32 \sigma}\psi~~~,
\end{equation}
\begin{equation}\label{2.32b}
\bar{H}_s^2=e^{-\sigma}H_s^2e^{\sigma}~~~,~~~
\bar{H}_d=e^{-\frac 32 \sigma}~ 
H_d ~e^{\frac 32 \sigma}~~~,
\end{equation}
where $\sigma=-\frac 12\ln |g_{00}|$. The inner products
for transformed functions 
can be
 obtained from Eqs. 
(\ref{product2}) and (\ref{product}) and they have  the 
universal form
\begin{equation}\label{product3}
(\bar{\Phi}_1,\bar{\Phi}_2)=\int_{\cal B}\sqrt{^{(3)}\bar{g}} 
~d^3x~
(\bar{\Phi}_1)^{+}\bar{\Phi}_2~~~.
\end{equation}
Here $\bar{\Phi}$ denotes either scalar or spinor field, 
$^{(3)}\bar{g}=\det \bar{g}_{ab}$, and
$\bar{g}_{ab}=g_{ab}/|g_{00}|=e^{2\sigma}g_{ab}$. Let us stress that
transformations (\ref{2.32}) and (\ref{2.32b}) do not change the
spectra determined by Eqs. (\ref{2.2'}). Thus, the  operators
$\bar{H}_i$ and $H_i$ ($i=s,d$) are {\it equivalent}.

From Eqs. (\ref{2.32b}) one finds the spinor
Hamiltonian
\begin{equation}\label{2.32bc}
\bar{H}_d=i\bar{\gamma_0}(\bar{\gamma}^a
\bar{\nabla}_a+e^{-\sigma}m )~~~,
\end{equation}
where $\{\bar{\gamma}^{\mu},\bar{\gamma}^{\nu}\}
=2\bar{g}^{\mu\nu}$ and $\bar{g}_{\mu\nu}=g_{\mu\nu}/|g_{00}|$. Thus,
for the square of scalar and spinor operators 
we have
\begin{equation}\label{2.33}
\bar{H}_i^2=-\bar{\nabla}^a
\bar{\nabla}_a+e^{-2\sigma}m^2+V_i~~~,~~~i=s,d~~~.
\end{equation}
The derivatives $\bar{\nabla}_a$ are defined in terms
of metric $\bar{g}_{ab}$
on the
three-dimensional hypersurface $\bar{\cal B}$.
The 
potential  $V_i$ is determined by the 
geometry of the background spaces and by 
the acceleration $w^\mu$ of the Killing observer
\begin{equation}\label{2.34a}
V_s=\xi\bar{R}+
e^{-2\sigma}(1-6\xi)
\left(\nabla^\mu w_\mu-w^\mu w_\mu\right)~~~,
\end{equation}
\begin{equation}\label{2.34b}
V_d=\frac 14 \bar{R}+e^{-2\sigma}m \gamma^\mu w_\mu~~~,
\end{equation}
\begin{equation}\label{2.34c}
\bar{R}=e^{-2\sigma}\left[R+
6(\nabla^\mu w_\mu-w^\mu w_\mu)\right]
~~~.
\end{equation}

Note that the operators $\bar{H}_i$ can be found if 
transformation (\ref{2.32}) of the fields is
applied to wave equations (\ref{2.18})--(\ref{2.19}).
The new equations obtained in this way can be 
expressed as equations of the fields on 
the {\it ultrastatic} spacetime with the 
metric
\begin{equation}\label{2.35a}
d\bar{s}^2=-dt^2+\bar{g}_{ab}dx^adx^b~~~,
\end{equation}
related to the physical metric by the conformal transformation, 
$\bar{g}_{\mu\nu}= g_{\mu\nu}/|g_{00}|$. The scalar curvature $\bar{R}$
for this metric is (\ref{2.34c}). Then developing the $(3+1)$-formalism
gives single-particle Hamiltonians (\ref{2.33}). In what follows all
quantities calculated with respect to the ultrastatic metric
$\bar{g}_{\mu\nu}$ will be denoted with a bar. The single-particle
spectra and canonical formulations of the  conformally related
theories  are equivalent. It should be emphasized that in the general
case (in the presence of mass $m$ and non-minimal coupling $\xi\neq
1/6$) the theory is not conformal invariant.  For scalar fields the
conformal invariance occurs only in the case when $m=0$ and $\xi=1/6$.

\subsection{Covariant Euclidean formulation}

The canonical formulation of 
statistical-mechanics is given in accordance
with the unitary evolution of the system
along the Killing time. It is 
explicitly related to the ``$3+1$'' decomposition
and, therefore, it is not explicitly covariant. 
The covariant Euclidean approach to quantum fields
at finite temperatures on stationary backgrounds
was suggested by Gibbons and Hawking 
\cite{GiHa:76}, \cite{Hawk:79}. This approach
proved to be especially useful in application
to thermodynamics of black holes 
\cite{GiHa:76}-\cite{BBWY:90}.

Consider a manifold ${\cal M}_\beta$ with the Euclidean metric 
\begin{equation}\label{3.2.1}
ds^2=g_{\tau\tau}d\tau^2 + 
g_{\alpha\beta}dx^\alpha dx^\beta~~,~~0\leq\tau\leq \beta~~,
\end{equation}
which is  obtained  from the static
Lorentzian metric (\ref{2.13}) by the Wick rotation $t\rightarrow
\tau=it$,  $g_{\tau\tau}=|g_{00}|$, and imposing the periodicity 
condition on the imaginary time $\tau$. We assume that the space is
asymptotically flat and has the topology $\R^3$, so that the topology of
${\cal M}_\beta$ is $\R^3\times S^1$. 

According to Gibbons and Hawking   \cite{GiHa:76}, \cite{Hawk:79}, the
partition function $Z^E$  and the effective action $W$  for a canonical
ensemble of the fields $\Phi$  in an external static  gravitational
background are defined by the path integral
\begin{equation}\label{3.1.1}
Z^E(\beta)=e^{-W[g,\beta]}=\int [D\Phi] e^{-I[g,\Phi]}~~~.
\end{equation}
Here $I[g,\Phi]$ is a classical Euclidean action on the manifold ${\cal
M}_\beta$.  Since $g_{\tau\tau}=1$ at spatial infinity, the parameter
$\beta$  is the length of $S^1$ and hence it has the meaning of the
inverse temperature measured at the spatial infinity. Fields $\Phi$ can
have either Bose or Fermi statistics. Bose variables are assumed to be
periodic in Euclidean time $\tau$ with the period $\beta$, while Fermi
fields are antiperiodic. $[D\Phi]$ is a covariant integration measure. 
For free scalar and Dirac fields the integration in (\ref{3.1.1}) gives
\begin{equation}\label{3.1.4}
W[g,\beta]=W_s[g,\beta]+W_d[g,\beta]~~~,
\end{equation}
\begin{equation}\label{3.1.6}
W_s[g,\beta]=\frac 12 \log\det\varrho ^{-2}L_s~~~,~~~
W_d[g,\beta]=-\log\det \varrho^{-1}L_d~~~.
\end{equation}
The functionals $W_i[g,\beta]$ are ultraviolet divergent and it is
assumed that their divergencies are regularized. $\varrho$ is an
arbitrary renormalization parameter with the dimension of the length,
which does not depend on the background metric. In what follows we  put
$\varrho=1$ for simplicity. If necessary the dependence  of the
effective actions on $\varrho$ can be easily restored. The operators
$L_i$ correspond to Eqs. (\ref{2.18}) and (\ref{2.19}) and read
\begin{equation}\label{3.1.7}
L_s=-\nabla^\mu\nabla_\mu+\xi R +m^2~~~,~~~
L_d=\gamma_5(\gamma^\mu\nabla_\mu+m)~~~.
\end{equation}
Note that under the Wick rotation from Lorentzian to Euclidean 
signature
the matrix $\gamma_0$ has to
be replaced by $i\gamma_0$. The matrix $\gamma_5$ 
anticommutes with other $\gamma$'s and is 
normalized
as $\gamma_5^2=1$. Both operators (\ref{3.1.7}) 
are Hermitean with respect to the standard
inner product
\begin{equation}\label{3.1.7ab}
\langle \Phi_1,\Phi_2\rangle=\int d^4x \,\sqrt{g}\, 
\Phi^{+}_1\Phi_2~~~. 
\end{equation}

The {\it Euclidean} free energy $F^E_i[g,\beta]$ is
defined by the effective action of the system
\begin{equation}\label{3.1.7a}
F^E_i[g,\beta]=\beta^{-1} W_i[g,\beta]-E^0_i[g]~~~.
\end{equation}
Similarly to $F^C_i[g,\beta]$, the so-defined 
Euclidean free-energy $F^E_i[g,\beta]$
vanishes at the zero temperature. This makes the
comparison of the two free energies more simple.
The vacuum energy
\begin{equation}\label{3.1.7b}
E^0_i[g]=\lim_{\beta\rightarrow \infty} 
\left(\beta^{-1}W_i[g,\beta]\right)~~~
\end{equation}
does not contribute to the entropy.
Since
the  free energy $F^E_i[g,\beta]$ and the vacuum energy $E^0_i[g]$
are defined in terms of the covariant Euclidean action 
$W_i[g,\beta]$ this approach is called {\em covariant Euclidean
formulation}.

\subsection{Relation between canonical and 
covariant Euclidean formulations}

The relation between  canonical and Euclidean formulations in the
absence of the horizon was discussed  by a number of authors
\cite{Gibbons:77}--\cite{DoKe:78}, \cite{Allen:86},
\cite{Dowker:84}--\cite{DoSc:89}.  

For the comparison of these two formulations the crucial role is played
by the representation of the canonical free energy in terms of the
effective action in the ultrastatic space $\bar{\cal M}_\beta$ with the
metric
\begin {equation}\label{3.2.2}
d\bar{s}^2=d\tau^2 + \bar{g}_{\alpha\beta} dx^\alpha dx^\beta~~~,
~~~0 \leq \tau \leq \beta~~~, 
\end {equation}
which is conformally related to ${\cal M}_\beta$.
This space is the product $S^1\times \bar{\cal B}$.

Let us define the operators $\bar{L}_i$ on 
$\bar{\cal M}_\beta$ which are conformally related
to operators $L_i$, Eq. (\ref{3.1.7}),
\begin{equation}\label{3.2.3}
\bar{L}_s=
e^{-3\sigma}L_s~e^{\sigma}~~~,~~~
\bar{L}_d=e^{-\frac 52 \sigma}L_d~e^{\frac 32 \sigma}~~~.
\end{equation}
It is easy to show that
\begin{equation}\label{3.2.4}
\bar{L}_s=\bar{H}_s^2-\partial^2_\tau~~~,
\end{equation}
\begin{equation}\label{3.2.5}
\bar{L}_d=\gamma_5\bar{\gamma}_\tau
(\bar{H}_d+\partial_\tau)~~~,~~~~
\bar{L}_d^2=\bar{H}_d^2-\partial^2_\tau~~~.
\end{equation}
For these operators one can define the effective actions
\begin{equation}\label{3.2.6}
\bar{W}_s[g,\beta]=\frac 12 \log\det\bar{L}_s~~~,~~~
\bar{W}_d[g,\beta]=-\log\det\bar{L}_d~~~.
\end{equation}
(For convenience we consider $\bar{W}_d$ as the functional
of the physical metric $g_{\mu\nu}$.)

For the system with a discrete spectrum one can  find with the help of
Eqs. (\ref{3.2.4}) and (\ref{3.2.5})  the 
following relation between the
canonical free energy and the effective action in the ultrastatic space
is \cite{Allen:86}
\begin{equation}\label{3.2.7}
F^C_i[g,\beta]=\beta^{-1} \bar{W}_i[g,\beta]-\bar{E}^0_i[g]
~~~.
\end{equation}
The quantity $\bar{E}^0_i[g]$
is the vacuum energy for the fields in the ultrastatic
space
\begin{equation}\label{3.2.8}
\bar{E}^0_i[g]=\eta_i\sum_\omega d_i(\omega) {\omega \over 2}~~~.
\end{equation}
Eqs. (\ref{3.2.7}) and (\ref{3.2.8}) can be generalized to include the
systems with continuous spectra.  For this purpose one has to take the
limit when the intervals between the levels vanish and to replace  sums
over $\omega$ with integrals. The derivation of (\ref{3.2.7}) and
(\ref{3.2.8}) can be found in work by  Allen \cite{Allen:86}, see also
\cite{Fursaev:97}.

It is important to note that $\bar{E}^0_i[g]$ includes all ultraviolet
divergences  of the functionals $\bar{W}_i[g,\beta]$.   Thus the
functional $F^C_i[g,\beta]$ is ultraviolet finite. The geometrical
structure of the divergences   does not depend on the temperature of 
the system \cite{DoKe:78}.  This is a consequence of a more general
property  of the quantum field theory whose ultraviolet singularities
do not depend on the  quantum state of the system \cite{BiDa:82}. The
renormalization of $\bar{W}_i[g,\beta]$  is equivalent to
renormalization of $\bar{E}^0_i[g]$.

Equations (\ref{3.2.3}) are crucial for finding relation between
canonical and Euclidean  free energies. The classical  actions
corresponding to the operators $L_i$ and $\bar{L}_i$ are
\begin{equation}\label{3.2.11}
I_i^E[g,\varphi_i]=\int_{{\cal M}_\beta}d^4x \,\sqrt{g}\,
\varphi^{+}_i L_i\varphi_i~~~,~~~
I^C_i[\bar{g},\bar{\varphi}_i]=\int_{\bar{{\cal M}}_\beta} 
d^4x\,\sqrt{\bar{g}}\,
\bar{\varphi}^{+}_i \bar{L}_i\bar{\varphi}_i
~~~,
\end{equation}
where the notation $\varphi_i$ is used for scalars $\phi$
or spinors $\psi$. So, as a result of (\ref{3.2.3}),
\begin{equation}\label{3.2.12}
I_i^E[g,\varphi_i]=I^C_i[\bar{g},\bar{\varphi}_i]
\end{equation} 
for $\bar{\phi}=e^{-\sigma}\phi$ and $\bar{\psi}=e^{-\frac 32
\sigma}\psi$. In case of massless scalars with
$\xi=\frac 16$ or massless spinors fields the operators 
$L_i$ and $\bar{L}_i$ have the same form
which means that classical theories  are conformally invariant. In
general case this invariance is broken. However, as it was shown by
Dowker and Schofield \cite{DoSc:88},\cite{DoSc:89} it is still possible
to introduce an auxiliary conformal charge in the classical actions
and    interpret Eq.(\ref{3.2.12}) in terms of a  pseudo conformal
invariance.

In quantum theory the classical symmetries are known to be broken
because of anomalies  \cite{BiDa:82},\cite{Duff:77}. Thus, the relation
between the renormalized actions has the form
\begin{equation}\label{3.2.9}
W_i[g,\beta]=\bar{W}_i[g,\beta]+
\beta\Omega_i[g]~~~.
\end{equation}
For scalar and spinors
fields the anomalous terms $\beta\Omega_i[g]$
were found explicitly in
\cite{DoSc:88},\cite{DoSc:89}.
The anomaly 
is proportional to $\beta$ and so it contributes
to the vacuum energy only 
\begin{equation}\label{3.2.10}
E^0_i[g]=\bar{E}^0_i[g]+\Omega_i[g]~~~.
\end{equation}
As a result, 
the free energies $F_i^E$ and $F_i^C$,
coincide 
\begin{equation}\label{3.2.10a}
F_i^E[g,\beta]=F_i^C[g,\beta]~~~.
\end{equation}
Therefore, on the static spacetimes without horizons the covariant
Euclidean and canonical formulations of statistical mechanics of
quantum fields are equivalent.

The appearance of the anomalous terms in (\ref{3.2.9}) can be
attributed to non invariance of the integration  measure  with respect
to conformal transformations \cite{Fujikawa:80}--\cite{Fujikawa:81}.
Path integral definition (\ref{3.1.1}) of functionals  $W_i[g,\beta]$
employs  the covariant measures  $g^{1/4}d\phi$ for scalars and
$g^{-r_d/2}d\psi^{+}d\psi$ for spinors,  where $g=\det g_{\mu\nu}$, and
$r_d$ is the dimensionality of the spinor representation (see for
instance Refs.\cite{FrVi:73}--\cite{Hawking:77}).  The difference in
the prefactors in scalar and spinor measures is related to the
different integration rules for Bose and Fermi variables. The
integration measure for actions  $\bar{W}_i[g,\beta]$ is different. It
is $\bar{g}^{1/4}d\phi$ and $\bar{g}^{-r_d/2} d\psi^{+}d\psi$, for
scalars and spinors, respectively, and it is covariant with respect  to
the ultrastatic background. One can obtain this form of the measure by
canonically  quantizing the theory \cite{deAOh:95}.

\section{Features related to the horizon}
\setcounter{equation}0

Many aspects of the computation of the statistical-mechanical
entropy of a thermal atmosphere of a black hole are directly related to
the presence of the horizon. Consider a general static spacetime with the
Killing vector $\zeta=\partial_t$ which has a Killing horizon where
$\zeta^2=0$. At the moment we do
not require that the metric obeys the Einstein equations.

If the surface gravity 
\begin{equation}
\kappa=\left[-{1\over 2}(\zeta_{\mu;\nu}
\zeta^{\mu;\nu})|_{\zeta^2=0}\right]^{1/2} 
\end{equation}
does not vanish,  the metric  near the horizon   can be represented in
the form
\begin{equation}\label{2.55}
ds^2\simeq -\kappa^2\rho^2 dt^2+d\rho^2 +d\Omega^2~~~.
\end{equation}
Here $d\Omega^2$ is the metric on  the two-dimensional bifurcation
surface of the horizons where $\zeta^\mu=0$. We denote this surface 
$\Sigma$. In coordinates (\ref{2.55}) the horizon is located at
$\rho=0$.   Such form of the metric is general for  non-extremal black
holes.

In the presence of the horizon the spectrum of quantum fields with
respect to the Killing time has a number of important new properties:
\begin{enumerate}
\item The single-particle spectrum is continuous 
even if the system has a finite size, i.e., when 
there are boundary conditions imposed on
the system at the finite distance from the horizon;
\item  Regardless of the spin and mass of the fields,
the spectrum of $\omega$ runs down to $\omega=0$
and, thus, the usual
mass gap is absent;
\item The bifurcation surface $\Sigma$ is invariant under
the time-evolution. 
\end{enumerate}

The continuity of the spectrum can be easily understood 
in representation  (\ref{2.33})
for the single-particle Hamiltonians.
The operators $\bar{H}_i$ are given on the spatial
part $\bar{\cal B}$ of the ultrastatic space (\ref{2.35a}).
The space $\bar{\cal B}$ is always
non-compact because the conformal transformation is singular at the
event horizon and it
moves the points of 
$\Sigma$ to the spatial infinity in $\bar{\cal B}$.

The mass gap for the operators $\bar{H}_i$ vanishes for the following
reason.  Equation (\ref{2.33}) shows that the mass $m_i$
of the field does not have any effect near the horizon because of the
factor $e^{-2\sigma}=|g_{00}|$. The operator
$-\bar{\nabla}^a\bar{\nabla}_a$ has  a mass gap which is of the pure
geometrical origin.  It arises since the space $\bar{\cal B}$ has
asymptotic constant curvature near the horizon.  However, on the
horizon  $V_s= -\kappa^2$, $V_d= -\frac 32\kappa^2$. These potentials
act as a tachionic mass \cite{CVZ:95a},\cite{BCZ:96}   which exactly
cancels the mass gap caused by the curvature of $\bar{\cal B}$.

Thus, the system we are dealing with behaves near the horizon similar
to a massless quantum theory on a non-compact space. It is well
known that such theories  run into the difficulties related to the infrared
divergences. Because of these divergences the densities of eigen-values
of quantum-mechanical Hamiltonians $\bar{H}_i$ blow up near the
horizon. As a result,  free energy (\ref{2.1ab})
in the canonical formulation diverges at any temperature.

\bigskip

In the presence of the horizon the covariant  Euclidean formulation
also exhibits new features. In this case the Euclidean section ${\cal
M}_\beta$, Eq. (\ref{3.2.1}), of the black hole solution
cannot be
regular for  arbitrary values of $\beta$. Near the surface $\Sigma$
where the Killing vector $\zeta=\partial_\tau$ vanishes 
(the {\em Euclidean horizon}) the metric
has the form 
\begin{equation}\label{3.4.1}
ds^2\simeq \kappa^2\rho^2 d\tau^2+d\rho^2 +d\Omega^2~~~,
~~~0\leq \tau \leq \beta~~~,
\end{equation}
which follows from Eq. (\ref{2.55}). Thus, near this surface ${\cal
M}_\beta$ looks as ${\cal C}_\beta\times  \Sigma$, where ${\cal
C}_\beta$ is a conical space.  It is easy to see that the conical
singularity  disappears and ${\cal M}_\beta$ is regular space only at
the special value   $\beta=\beta_H=2\pi\kappa^{-1}$.  The
corresponding temperature $T_H=\beta_H^{-1}$ coincides with the Hawking
temperature and the corresponding quantum state is  known as the
Hartle-Hawking vacuum \cite{HaHa:76}.  The physical meaning of $T_H$
is  that it is the temperature of the Hawking quanta  emitted by an
evaporating black hole.  $T_H$ also gives the temperature at which the 
quantum radiation can be in thermal equilibrium with a black hole. 

For $\beta\neq \beta_H$ conical singularities result in additional
ultraviolet divergences in the effective action on ${\cal M}_\beta$.
Thus,  in the presence of the horizon both canonical and covariant
Euclidean formulations acquire new divergences. These divergences are
of the different origin; they are infrared in the canonical method, and
ultraviolet in the covariant Euclidean approach. In  Sections 5 and 6
we develop canonical and covariant Euclidean formulation in the
presence of the horizon and  establish the relation between the
divergences in these formulations.

\section{Canonical formulation in the presence of a horizon}
\setcounter{equation}0

\subsection{Density of levels and its properties}

The divergence of statistical-mechanical  quantities in the presence
of a horizon is directly related to the infinite growth of the
density of states  of the Hamiltonians. To investigate this property we
follow Ref. \cite{Fursaev:97}. The idea, which is close
to the earlier approach of \cite{CVZ:95a}, is
to relate ${dn_i(\omega)/d\omega}$  to the heat kernel of the operator
$\bar{H}_i^2$.  The latter is an elliptic operator and its heat kernel
is well known. For a continuous spectrum one has
\begin{equation}\label{2.4.1}
\mbox{Tr}~e^{-\bar{H}_i^2t}=\int_{0}^{\infty}
d\omega {dn_i(\omega) \over d\omega} e^{-\omega^2t}~~~.
\end{equation}
The density ${dn_i(\omega) / d\omega}$ can be found from (\ref{2.4.1})
with the help of the inverse Laplace transform  \cite{Bateman:54}. 
The diagonal matrix elements 
$\langle x|\exp(-\bar{H}_i^2t)|x \rangle\equiv
\left[\exp(-\bar{H}_i^2t)
\right]_{\mbox{diag}}$ 
are well defined and finite. However, the corresponding  trace which
involves the integration over the noncompact space $\bar{\cal B}$
diverges. To understand why it happens it is sufficient to study the
behavior of  $\left[\exp(-\bar{H}_i^2t) \right]_{\mbox{diag}}$ near the
horizon. To estimate the leading asymptotics  we can neglect the
curvature of the two dimensional surface of the horizon and 
approximate the black hole
metric (\ref{2.55}) by the metric on the Rindler space
\begin{equation}\label{rind}
ds^2= -\kappa^2\rho^2 dt^2+d\rho^2+dz_1^2+dz_2^2
~~~,~~~
-\infty < z_1,z_2 <\infty~~~,~~~\rho > 0~~~.
\end{equation}
Then the metric 
on the conformal space $\bar{\cal B}$  
\begin{equation}\label{2.4.3}
dl^2=\kappa^{-2}\rho^{-2}(d\rho^2+dz_1^2+dz_2^2)~~~
\end{equation}
coincides with the metric of the hyperbolic manifold $\ag^3$ of
constant negative curvature $\bar{R}=-6\kappa^2$. A 
review  of the Laplace operators  and their heat kernels on  
hyperbolic spaces can be found in Refs. \cite{Camporesi:90} and
\cite{BCVZ:96}.  The eigen-functions of $\bar{H}_i^2$ are completely
determined by the requirement to have the correct decay properties at
infinity $\rho\rightarrow\infty$ \cite{BCZ:96}. Thus, no additional
conditions  at the horizon $\rho=0$ are needed. When $\rho\rightarrow
0$ the fields become  effectively massless and the  diagonal elements
of the scalar and spinor heat kernels on $\ag^3$ are known exactly 
\cite{BCZ:96},\cite{BCVZ:96}
\begin{equation}\label{2.4.4}
\left[ e^{-t\bar{H}_s^2}\right]_{\mbox{diag}}
= {1 \over (4\pi t)^{3/2}}
~~~,
~~~
\left[ e^{-t\bar{H}_d^2}\right]_{\mbox{diag}}
=  {r_d \over (4\pi t)^{3/2}}
\left(1+\frac 12 \kappa^2 t\right)~~~.
\end{equation}
A summation over the spinor indexes is assumed and  it gives the factor
$r_d$. The geometry  essentially differs from the Rindler one far from
the horizon and  the mass term becomes important. For this reason in
general there are corrections to Eqs. (\ref{2.4.4}) proportional to the
powers of $\rho^2$. As it was shown in \cite{Fursaev:97}, the
structure  of these terms can be analysed  by using  asymptotic
properties of the heat kernels on $\bar{\cal B}$.

Therefore, as follows from Eqs. (\ref{2.4.1}) and (\ref{2.4.4}), the
trace of the operators and their density of  states grows as the volume
of $\bar{\cal B}$. 

There are several ways how to regularize  these divergences. For
instance, one can restrict the spatial size of the system. In the given
case it means that  the region of the  physical spacetime where the
proper distance to the horizon is smaller than some length, say
$\epsilon$, has to be excluded from the consideration.  To this aim
t'Hooft \cite{Hooft:85} suggested to impose the  Dirichlet boundary
conditions on the fields at a surface located outside the horizon and
at the proper distance $\epsilon$ from it. t'Hooft's approach is known
as the {\em ``brick wall'' model}. A similar but  simpler procedure,
called the {\em volume cutoff} method,  was proposed by Frolov and
Novikov \cite{FrNo:93} (see also Refs. \cite{FFZ:96a},\cite{ZCV:96}).
In this method  all spatial integrations are cut off at the proper
distance $\epsilon$ without imposing a boundary  condition.

In the volume cutoff method one effectively cuts a region near the
horizon which makes the spacetime incomplete. There exist other types
of regularizations which allow one to work on the complete spacetime
background. An example is the Pauli-Villars (PV) regularization which
was first used for the problem by  Demers, Lafrance and  Myers
\cite{DLM:95}. The density ${dn/d\omega}$ turns out to be finite even
in the limit $\epsilon\rightarrow 0$  but it depends on the regulator
mass $\mu$ and for $\mu\rightarrow \infty$ it grows as $\mu^2$. Another
option suggested in \cite{FFZ:97} is to
use the dimensional regularization. The idea of this method is that the
power of the leading divergency in  Eqs. (\ref{2.4.5}) and
(\ref{2.4.6}) depends on the number of spacetime dimensions. In
$D$-dimensional spacetime the leading divergence of the volume of
$\bar{\cal B}$ is $\epsilon^{2-D}$, if $D\neq 2$. One can use $D$ as a
regularization parameter for density of states, and take the limit 
$\epsilon\rightarrow 0$ at $\mbox{Re} D< 2$.  The density
${dn/d\omega}$  then has a pole at $D=4$.

In the volume cutoff method the divergence of ${dn/d\omega}$ is
infrared.  In dimensional and PV regularizations the divergence of ${dn
/d\omega}$ can be directly connected with standard ultraviolet
divergences. For this reason, we  can speak about  infrared and ultraviolet
limits of the theory depending on which type of regularization is used.

\subsection{Infrared limit and volume cutoff}

Let us denote the regularized density  in the volume cutoff method as 
${dn_i(\omega |\epsilon)/d\omega}$ and investigate its asymptotics in
the limit $\epsilon\rightarrow 0$. After integrating the heat kernels
over the region $\rho\geq\epsilon$ and using the inverse Laplace
transform in (\ref{2.4.1})  one obtains the
regularized expression for the divergent part of 
densities of states \cite{Fursaev:97}
\begin{equation}\label{2.4.5}
\left[{dn_s(\omega |\epsilon) \over d\omega}\right]
_{\tiny\mbox{div}}
= {1 \over 4\pi^2 \kappa^3}\int_{\Sigma}
\left\{\omega ^2
\left({1 \over \epsilon^2}-\frac 14 {\cal P}
\ln {\epsilon^2 \over l^2}\right)
-{\kappa^2 \over 2}
\ln {\epsilon^2 \over l^2}
\left[\left(\frac 16-\xi\right) R-m^2\right]\right\}
~~~,
\end{equation}
$$
\left[{dn_d(\omega |\epsilon) \over d\omega}\right]
_{\tiny\mbox{div}}=r_d
{1 \over 4\pi^2 \kappa^3}\int_{\Sigma}
\left\{
\omega ^2
\left({1 \over \epsilon^2}-\frac 14 {\cal P}
\ln {\epsilon^2 \over l^2}\right)
+{\kappa^2 \over 4\epsilon^2}
\right.
$$
\begin{equation}\label{2.4.6}
\left.
-{\kappa^2 \over 2}\ln {\epsilon^2 \over l^2}
\left(\frac 18 {\cal Q}-{1 \over 12} R-m^2\right)
\right\}~~~.
\end{equation}
The notation $\int_{\Sigma}$ assumes that the integration is performed
over the bifurcation surface of the horizons $\Sigma$, so that
$\int_{\Sigma} 1={\cal A}$,  where ${\cal A}$ is the area of 
$\Sigma$.  We imposed  an additional cutoff $l$ at large distance $\rho$. The
quantity $r_d$ is the
dimension of the spinor representation, so that $r_d=4$ for 4
dimensional Dirac spinors. 
Let $n^{\mu}_i$ $(i=1,2)$ be two unit mutually orthogonal vectors normal to
$\Sigma$, then  $P^{\mu\nu}=\sum_{i=1}^2 n^{\mu}_i n^{\nu}_i$ is a projector
onto a two dimensional surface orthogonal to $\Sigma$. The quantities
${\cal P}$ and ${\cal Q}$ which enter (\ref{2.4.5}) and (\ref{2.4.6}) 
are defined as
\begin{equation}\label{2.6.3}
{\cal P}=2{\cal R}-{\cal Q}~~~,~~~{\cal Q}=
P^{\mu\nu}R_{\mu\nu}~~~,~~~
{\cal R}=P^{\mu\nu}P^{\lambda\rho}R_{\mu\lambda\nu\rho}~~~.
\end{equation}

Note that the leading divergence  $\epsilon^{-2}$ in Eqs.
(\ref{2.4.5})--(\ref{2.4.6})  is already present in the Rindler approximation
(\ref{2.4.4}). The mass and non-zero curvature result in  the additional
logarithmic divergences $\ln(\epsilon^2/l^2)$.

\bigskip

Substituting (\ref{2.4.5}) and (\ref{2.4.6}) into  expression
(\ref{2.1ab}) for the canonical free energy  we obtain 
\begin{equation}\label{2.5.1}
F^C_{s,\tiny\mbox{div}}[g,\beta,\epsilon]=
-{1 \over \kappa^3} \int_{\Sigma}
\left\{{\pi^2 \over 180  \beta^4\epsilon^2}
-
\left[{\pi^2 \over 720 \beta^4}{\cal P}
+
{\kappa^2 \over 48 \beta^2}
\left(\left(\frac 16-\xi\right) R-m^2\right)\right]
\ln {\epsilon^2 \over l^2}
\right\}~~~,
\end{equation}
$$
F^C_{d,\tiny\mbox{div}}[g,\beta,\epsilon]=
-r_d{1 \over \kappa^3}\int_{\Sigma}
\left\{\left({7\pi^2 \over 1440 \beta^4 }+
{\kappa^2 \over 192\beta^2}
\right){1 \over \epsilon^2}\right.
$$
\begin{equation}\label{2.5.2}
\left.
-\left[{7\pi^2 \over 5760 \beta^4 } {\cal P}
+{\kappa^2 \over 96 \beta^2}
\left(\frac 18 {\cal Q}-{1 \over 12} R-m^2\right)\right]
\ln {\epsilon^2 \over l^2}
\right\}~~~.
\end{equation}
This  enables one to calculate other  characteristics of canonical
ensembles. In particular, one can find the $\epsilon$ 
divergence of the  statistical-mechanical entropy
\begin{equation}\label{2.5.3}
S^C_{i,\tiny\mbox{div}}[g,\beta,\epsilon]=\beta^2 
{\partial F^C_{i,\tiny\mbox{div}}[g,\beta,\epsilon] 
\over \partial \beta}~~~.
\end{equation}
Evaluated at the Hawking temperature $\beta^{-1}= \beta_H^{-1}\equiv
\kappa / 2\pi$ the divergent part of 
the contributions of bosons and fermions to
the entropy is 
\begin{equation}\label{2.5.4a}
S^C_{s,\tiny\mbox{div}}[g,\beta_H,\epsilon]=
{1 \over \pi} \int_{\Sigma}
\left\{{1 \over 360 \epsilon^2}
-
{1 \over 1440}\left[2{\cal R}-{\cal Q}
+
30\left(\frac 16-\xi\right) R-30m^2\right]
\ln {\epsilon^2 \over l^2}
\right\}~~~,
\end{equation}
\begin{equation}\label{2.5.4b}
S^C_{d,\tiny\mbox{div}}[g,\beta_H,\epsilon]=
r_d{1 \over \pi}\int_{\Sigma}
\left\{{11 \over 1440 \epsilon^2}
-{1 \over 5760 }\left[7 {\cal R} +4 {\cal Q}-5 R
-60 m^2\right]
\ln {\epsilon^2 \over l^2}
\right\}~~~.
\end{equation}
As we already mentioned,  the vacuum energy omitted in  (\ref{2.24})
does not contribute to the entropy.  For scalars, the leading
divergence $\epsilon^{-2}$ of the entropy is determined by $\beta^{-4}$
term in free energy (\ref{2.5.1}). For
the spinor  fields the situation is different. In order to find
$\epsilon^{-2}$ divergence to the entropy one has to know both the
leading, $\beta^{-4}$, and subleading, $\beta^{-2}$, terms in free
energy (\ref{2.5.2}).

\bigskip

To determine the divergences of the free energy and entropy one has to
know only the asymptotics of the heat kernel operators
\cite{Fursaev:97}. This is not sufficient if one wants to calculate
not only the divergencies but the quantities themselves. Important
examples, when ${dn/d\omega}$, $F^C$ and $S^C$
 can be
calculated exactly, were studied in Refs.
\cite{CVZ:95a}, \cite{BCZ:96}.  The authors considered the case when
the spatial part of the spacetime is of  the form  ${\cal
B}=\R^{+}\times \Sigma$, where $\Sigma$ is a  manifold of
constant curvature.  In particular, for the Rindler spacetime $\Sigma$
is a two dimensional plane and the density of states of massive scalar
fields in  Rindler space is \cite{CVZ:95a}
\begin{equation}\label{2.57}
{dn_s(\omega |\epsilon) \over d\omega}=
{{\cal A} \over 4\pi^2\kappa^3}
\left[{\omega ^2 \over \epsilon^2}
+{m^2\kappa^2 \over 2}\ln{\epsilon^2m^2\over 4}
-{m^2\kappa^2 \over 2}
\left(1+2\mbox{Re}~\psi(i\omega/\kappa)\right)\right]
~~~.
\end{equation}
Here $\psi$ is the logarithmic derivative of the $\Gamma$-function. The
terms which vanish as $\epsilon\rightarrow 0$ are omitted. It follows
from (\ref{2.57}) that the  role of additional cutoff $l$ in
(\ref{2.4.5}) and (\ref{2.4.6}) is played by $m^{-1}$.

\subsection{Ultraviolet limit}

\noindent
{\bf A. Dimensional regularization}

\bigskip

\noindent
The dimensional regularization is the simplest scheme which enables one
to define ${dn/d\omega}$ on the complete spacetime. In this
regularization one can put $\epsilon=0$ from the very beginning. 
The quantity ${dn/d\omega}$  depends on the complex parameter $D$ 
associated with the dimensionality of the spacetime and it has a pole
singularity when $D=4$ \cite{Fursaev:97}
\begin{equation}\label{2.6.1}
\left[{dn_s(\omega|D) \over d\omega}\right]_{\tiny\mbox{div}}=
{\Gamma\left(1-\frac D2\right) \over (4\pi)^{D/2}}\,\,
 {m^{D-4} \over \kappa}\,\, \int_{\Sigma}
\left[2\left(m^2-\left(\frac 16 -\xi\right)R\right)
-{\omega^2\over \kappa^2}{\cal P}\right]~~~,
\end{equation}
\begin{equation}\label{2.6.2}
\left[{dn_d(\omega|D) \over d\omega}
\right]_{\tiny\mbox{div}}=
r_d\, {\Gamma\left(1-\frac D2\right) \over (4\pi)^{D/2}}
\,\, {m^{D-4} \over \kappa}\,\, \int_{\Sigma}
\left[2\left(m^2+{R\over 12} -{{\cal Q} \over 8}\right)
-{\omega^2\over \kappa^2}{\cal P}\right]~~~.
\end{equation}
These expressions can be used to get the divergences of
the canonical free energy for scalars and 
spinors
\begin{equation}\label{2.6.4}
F^C_{\tiny\mbox{div}}[g,\beta,D]=-\eta\,
{\Gamma\left(1-\frac D2\right) \over (4\pi)^{D/2}}
\,\,{\pi^2 m^{D-4} \over 3\kappa \beta^2}\,\,
\int_{\Sigma}
\left[f_1 m^2- \left(p_1
{4\pi^2 \over \kappa^2\beta^2} {\cal P}+
p_2 R+p_3 {\cal Q})\right)
\right]~~~.
\end{equation}
The corresponding divergence of the entropy is
$$
S^C_{i,\tiny\mbox{div}}[g,\beta,D]=
\beta^2\, {\partial F^C_{i,\tiny\mbox{div}}[g,\beta,D] 
\over \partial \beta}~~~,
$$
\begin{equation}\label{2.6.5}
S^C_{i,\tiny\mbox{div}}[g,\beta,D]=\eta\,
{\Gamma\left(1-\frac D2\right) \over (4\pi)^{D/2}}\,\,
{2\pi^2 m^{D-4} \over 3\kappa \beta}\,\,
\int_{\Sigma}
\left[f_1 m^2- \left(p_1
{8\pi^2 \over \kappa^2\beta^2} {\cal P}+
p_2 R+p_3 {\cal Q})\right)
\right]~~~.
\end{equation}

Constants $f_1$ and $p_k$ depend on spin and are given in Table 1. 
This table contains also other similar constants which we meet later. 

\bigskip

\begin{table}[t]
\renewcommand{\baselinestretch}{2}
\medskip
\caption{}
\bigskip
\begin{centerline}
{\small
\begin{tabular}{|c||c||c||c|c|c||c|c|c|}
\hline
$\mbox{spin}$  & $d_1$ & $f_1$    
& $q_1$   & $q_2$   & $q_3$ 
& $p_1$   & $p_2$   & $p_3$   \\
\hline
\hline
$0$ &  $\frac 16 -\xi$ & $1$ & 
$1$ & $-1$  & $\frac 52(1-6\xi)^2$  &
${1 \over 60}$ & $\frac 16-\xi$  & $0$  \\
\hline
\hline
$\frac 12$ & $-{1 \over 12} r_d$ & $-\frac 12 r_d$ & 
$-{7 \over 8}r_d$ & $-r_d$  & 
${5 \over 8}r_d$ &
$-{7 \over 480}r_d$ & ${1 \over 24}r_d$  & 
$-{1 \over 16}r_d$  \\
\hline
\end{tabular}}
\bigskip
\renewcommand{\baselinestretch}{1}
\end{centerline}

\end{table}

\bigskip
\bigskip

\noindent
{\bf B. Pauli-Villars regularization}

\bigskip

\noindent
It is known that the dimensional regularization  reproduces  only the
logarithmic divergences.  For this reason it is also worth
studying $F^C_{\tiny\mbox{div}}$ in another, more complete  regularization.
For our purpose it is convenient to  use the {\em Pauli-Villars method}. In
this method for each of the physical field, one introduces   $5$ 
additional 
auxiliary fields: $2$ fields with
masses $M_k$ which have the same statistics as the original field and  $3$
fields with masses $M_r'$ which have the wrong statistics\footnote{They
are fermions for scalars and bosons for spinors.}.  To eliminate the
divergences the masses  of the auxiliary fields must obey the two 
restrictions 
\begin{equation}\label{3.4}
f(1)=f(2)=0~~~,
\end{equation}
where
\begin{equation}\label{3.4ab}
f(p)=m^{2p}+\sum_k M_k^{2p}-\sum_r(M'_r)^{2p}=0~~~.
\end{equation}
These constraints can be resolved by 
taking 
$M_{1,2}=\sqrt{3\mu^2+m^2}$, $M'_{1,2}=\sqrt{\mu^2+m^2}$,
$M'_3=\sqrt{4\mu^2+m^2}$ (see \cite{DLM:95}).
The regularized density of states in this method
is 
\begin{equation}\label{3.4a}
{dn_i(\omega|\mu) \over d\omega}\equiv
{dn_i(\omega, m) \over d\omega}+\sum_k 
{dn_i(\omega,M_k) \over d\omega}
-\sum_r {dn_i(\omega,M'_r) \over d\omega}~~~.
\end{equation}
The quantities ${dn_i(\omega,M_k)/d\omega}$ and 
${dn_i(\omega,M_r')/d\omega}$ are the density of states of the
Pauli-Villars partners. The fields with the wrong statistics give
negative contribution in the regularized density. Because the number of
such fields equals the number of the fields with the proper statistics
the leading  $\epsilon$-divergences in relations (\ref{2.4.5}) and
(\ref{2.4.6}) are cancelled. Logarithmic  divergences $\ln\epsilon^2$
are also canceled because of the constraint $f(p=1)=0$ 
imposed on masses. As a result,  the regularized density of
states (\ref{3.4a}) does not contain the divergences when
$\epsilon\rightarrow 0$ and it can be defined on the complete
background. Since in the presence of regulators ${dn_i(\omega|\mu) /
d\omega}$ is finite one obtains the same answer as when
one uses the dimensional
regularization for its calculation. Using 
Eqs. (\ref{2.6.1}) and (\ref{2.6.2}) and by taking into account constraints
(\ref{3.4}) one gets
\begin{equation}\label{33.1}
\left[{dn_s(\omega |\mu) \over d\omega}\right]_{\tiny\mbox{div}}=
{1 \over (4\pi)^2 \kappa}
 \int_{\Sigma}
\left[2b+a\left({\omega^2 \over \kappa^2}{\cal P}+
2\left(\frac 16 -\xi\right)R\right)\right]~~~,
\end{equation}
\begin{equation}\label{33.2}
\left[{dn_d(\omega |\mu) \over d\omega}
\right]_{\tiny\mbox{div}}=
r_d {1 \over (4\pi)^2 \kappa}
\int_{\Sigma}
\left[2b+a \left({\omega^2 \over \kappa^2}{\cal P}
-{R\over 6} +{{\cal Q} \over 4}\right)
\right]~~~.
\end{equation}
The functions $a$ and $b$ depend on $m$ and $\mu$
\begin{equation}\label{3.5b}
a\equiv -\left.{df\over dp}\right|_{p=0}=-
\ln m^2-\sum_k \ln M_k^2+ 
\sum_r \ln (M'_r)^2~~~,
\end{equation}
\begin{equation}\label{3.5c}
b\equiv \left.{df\over dp}\right|_{p=1}=
m^2\ln m^2+\sum_k  M^2_k\ln M_k^2- 
\sum_r (M'_r)^2\ln (M'_r)^2~~~.
\end{equation}
In the Pauli-Villars method $\mu^2$ plays the role of the ultraviolet
cutoff. In the  limit of infinite masses $M_k$, $M'_r$,
i.e. at $\mu\rightarrow \infty$,  ${dn_i(\omega|\mu)/d\omega}$
is ultraviolet divergent.  In
this limit 
\begin{equation}\label{3.5d} 
a\simeq \ln {\mu^2 \over
m^2}~~~,~~~ b\simeq \mu^2\ln{729 \over 256}-m^2\ln{\mu^2 \over m^2}~~~.
\end{equation} 
Thus, in general ${dn/d\omega}$ includes  both   quadratic and
logarithmic divergences.

From Eqs.  (\ref{33.1})--(\ref{33.2}) one can derive the divergence of
the statistical-mechanical free energy
\begin{equation}\label{3.5}
F^C_{\tiny\mbox{div}}[g,\beta,\mu]=-
{\eta \over 48\kappa \beta^2}
\int_{\Sigma}
\left[b f_1 + a \left(p_1
{4\pi^2 \over \kappa^2\beta^2} {\cal P} +
p_2 R+p_3 {\cal Q}\right)
\right]~~~.
\end{equation}
A special case of expression (\ref{3.5}) for scalar fields on 
the Reissner-Nordstr\"om
black hole background was first derived in
\cite{DLM:95}.  The
divergent part of the entropy obtained from  (\ref{3.5}) at 
the Hawking temperature is
\begin{equation}\label{3.6a}
S^C_{\tiny\mbox{div}}[g,\beta_H,\mu]=
{\eta \over 48\pi}
\int_{\Sigma}
\left[b f_1 + a \left(2p_1{\cal P} +
p_2 R+p_3 {\cal Q}\right)
\right]~~~.
\end{equation}
By taking into account Eq. (\ref{3.5d}) we can rewrite this expression
at large $\mu$ in the form
\begin{equation}\label{3.6b}
S^C_{\tiny\mbox{div}}[g,\beta_H,\mu]=
{\eta \over 48\pi}
\int_{\Sigma}
\left[c \mu^2  f_1 + \left(2p_1{\cal P} +
p_2 R+p_3 {\cal Q}-f_1 m^2\right)\ln{\mu^2 \over m^2}
\right]~~~,
\end{equation}
where $c=\ln{729 \over 256}$. Expression (\ref{3.6b}) has 
the same structure   as the entropy divergences  (\ref{2.5.4a}) and
(\ref{2.5.4b}) in the volume cutoff regularization. It is easy to see
that in these regularizations parameters $\epsilon$  and $\mu^{-1}$
correspond to each other.  
The fact that Pauli-Villars regularization  results in a cutoff of the
integrals near the horizon at the proper distance comparable to the
inverse  masses of the  fields allows the following interpretation. Near
the horizon where the local temperature becomes greater than $\mu$, the
massive regulators are thermally excited and because of constraints
(\ref{3.4}) their contribution exactly cancels the contribution of the
physical field.

It is interesting that for each field one can find  a relation
between $\epsilon$ and $\mu^{-1}$ by equating the leading 
divergencies. After that it is possible to find the connection between
$l$ with $m^{-1}$ which makes equal logarithmic divergencies as well.
Note, however, that in this identification the relation between
$\epsilon$ and $\mu$ is different for fields of the different
spins, see Eqs. (\ref{2.5.4a}), (\ref{2.5.4b}) and (\ref{3.6b}).

\subsection{WKB computations and the ``brick wall'' model}

The divergences in statistical-mechanical quantities  can be also
obtained by using WKB method. This way of computations was suggested by
t'Hooft \cite{Hooft:85} and then used by many   authors, see for
instance,  Refs. \cite{MTZ:92}--\cite{CoLe:97}. To illustrate this
method we consider a scalar field on asymptotically flat spherically
symmetric black-hole background
\begin{equation}\label{2.3.1}
ds^2=-g(r)dt^2+g^{-1}(r)dr^2+r^2(d\theta^2+
\sin^2\theta d\varphi^2)~~~.
\end{equation}
Here $r\geq r_h$, and $r_h$ is the horizon
radius where $g(r_h)=0$. Schwarzschild and 
Reissner-Nordstr\"om 
black holes are described by metrics of this type. 
We are interested in modes $\bar{\phi}_{\omega,\ell}({\bf x})$  with 
energy $\omega$ and angular momentum $\ell$.
They are solutions of the eigen-value problem 
\begin{equation}\label{2.3.2}
\bar{H}_s^2\bar{\phi}_{\omega,\ell}=
\omega^2\bar{\phi}_{\omega,\ell}~~~.
\end{equation}
This equation  is reduced to one-dimensional problem
\begin{equation}\label{2.3.3}
\left[-{g^2 \over r^2}\partial_r \left(r^2\partial_r\right)
+V^{\ell}_{eff}(r)-\omega^2\right]\bar{\phi}_{\omega,\ell}=0~~~
\end{equation}
with the potential
\begin{equation}\label{2.3.4}
V^{\ell}_{eff}(r)=-{1\over 4}(g')^2+g(r)\left[m^2+r^{-2}
\ell(\ell+1)\right]~~~.
\end{equation}
Near the horizon this potential is negative
$V^{\ell}_{eff}(r_h)=
-\kappa^2$, while at $r\gg r_h$
it is positive and $V^{\ell}_{eff}(r=\infty)=m^2$.

Let us consider now the "brick wall" model
\cite{Hooft:85}.  
We assume that the ``brick wall'' Dirichlet condition is 
imposed at the proper distance $\epsilon$ near the horizon, and
$r(\epsilon)$ is the location of the "brick wall" in coordinates 
(\ref{2.3.1}). 
Equation (\ref{2.3.3})
enables one to estimate in the quasiclassical approximation the number
of energy levels $n_s(\omega |\epsilon)$ with the energy smaller than 
$\omega$  
\begin{equation}\label{2.3.5}
n_s(\omega |\epsilon)={1 \over \pi}
\sum_{\ell=0}^{\ell(\omega)}(2\ell+1)
\int_{r(\epsilon)}^{r_B}
{dr \over g(r)}\sqrt{\omega^2-V^{\ell}_{eff}(r)}~~~.
\end{equation}
Here $\ell(\omega)$ is the maximal angular momentum 
at which the square root in (\ref{2.3.5}) vanishes, 
$r_B\gg r_h$ is an additional infrared cutoff.

To estimate the asymptotic behavior of $n_s(\omega |\epsilon)$ at
$\epsilon\rightarrow 0$, we assume that  $\omega^2$ is large compared to
the curvature of  the background. Then only the contribution  of the
large momenta is important and the sum over $\ell$ can be replaced by the
integral, which can be easily calculated
\begin{equation}\label{2.3.6}
n_s(\omega |\epsilon)\simeq {2 \over 3\pi}
\int_{r(\epsilon)}^{r_B}
{r^2 dr \over g^2}(\omega^2-g m^2)^{3/2}
\simeq {{\cal A} \over 12\pi^2\kappa^3}
\left({\omega^3 \over \epsilon^2}+\frac 32 \kappa^2 m^2 
\omega \ln {\epsilon^2 \over l^2}\right)~~~.
\end{equation}
In the last equality we put ${\cal A}=4\pi r_h^2$, $d\rho=g^{-1/2}dr$
and use 
for the metric (\ref{2.3.1}) near the
horizon the Rindler approximation (\ref{rind}). 
It is easy to see that Eq.(\ref{2.3.6}) gives the same 
density of energy levels as expression (\ref{2.4.5}) for a  scalar
field with ${\cal P}=R=0$.

Thus, at least for the leading divergences  there is an agreement
between the volume cutoff regularization and the WKB ``brick wall''
model. This agreement  remains if one chooses the Newman boundary
condition instead of the Dirichlet one. The numbers
$n_s(\omega |\epsilon)$ for these conditions differ only by a numerical
constant \cite{KaSt:94}. It should be noted that for the physical field
the horizon is not a real boundary.  From the mathematical point of
view it means that the wave functions in Eq.(\ref{2.3.2}) with
$\omega<m$ can be
completely fixed by the requirement to decay fast enough  at $r\gg r_h$
and no boundary conditions at the cutoff length $\epsilon$ are
required.  For this reason the volume cutoff method seems to be more
appropriate than the ``brick wall'' approach.

The WKB method can be also used to study the ultraviolet divergences. 
It was employed, for instance, in \cite{DLM:95} in PV regularization.
Analogous computations  were done in Refs.
\cite{Solod:96a} and \cite{KKSY:97}. Almost all WKB computations  are
restricted to the case of the  scalar field  (see, however,
Refs. \cite{CoLe:97} and \cite{SCZ:97}). 

\bigskip

\section{Covariant Euclidean formulation}
\setcounter{equation}0

\subsection{Volume cutoff and high-temperature expansion}

In covariant Euclidean formulation of the theory, the problems
related to the divergences on the horizon can be
treated in the same way as in the canonical formulation;
namely, by introducing either the volume cutoff 
or ultraviolet regularizations. 
Obviously, comparing the Euclidean and canonical
formulations has sense only for the equivalent 
regularizations. 

\bigskip

In the volume cutoff regularization, one effectively cuts  a region near
the horizon. This makes the spacetime incomplete. As the result, one
gets a theory  on a static spacetime  without the horizon. As we
discussed earlier,   the covariant Euclidean and canonical formulations
are equivalent in this case. 

The form of the Euclidean action  near the
horizon can be found by using the high temperature asymptotics. These
asymptotics give a good approximation  because the local temperature
infinitely grows when approaching the horizon. The high temperature 
expansion of an effective action was obtained by
Dowker and collaborators  in \cite{DoKe:78}, \cite{DoSc:88} and
\cite{DoSc:89}  (see also Ref. \cite{NaFu:85}). The
expansion has the form 
\begin{equation}\label{2.51}
W_i[g,\beta]=
\beta\int d^3x\,{g}^{1/2}\,\left[b_i(x,\beta)+h_i(x)\right]
+W_i^{(3)}[g]-\Delta W_i~~~.
\end{equation}
For  spinors $W_d^{(3)}[g]=0$, while for scalars
$W_s^{(3)}[g]=-\frac 12 \zeta'(0|\bar{H}^2_i)$ is  determined by the
$\zeta$-function of the operator $\bar{H}_s^2$, see
\cite{DoSc:88},\cite{DoSc:89}. The quantities $b_i$ and $h_i$ which
enter relation (\ref{2.51}) are
\begin{equation}\label{2.52}
b_s(x,\beta)=-{\pi^2 \over 90\beta_l^4}
-{1 \over 24 \beta_l^2}
\left[\left(\frac 16-\xi\right)R-m^2\right]
-{a_{s,2}(x) \over 16\pi^2}\ln {\mu\beta_l \over 2\pi}~~~,
\end{equation}
\begin{equation}\label{2.53}
b_d(x,\beta)=-{7\pi^2 r_d \over 720\beta_l^4}
+{r_d \over 48 \beta_l^2}\left[{1 \over 12} R+
\frac 12 (\nabla w -w^2)+m^2\right]
+{a_{d,2}(x) \over 16\pi^2}
\ln {\mu\beta_l \over 2\pi}~~~,
\end{equation}
\begin{equation}\label{2.52a}
h_s={1 \over 2880\pi^2}\left[5w^2(w^2-2\nabla w)
-3(\nabla w)^2+R_{\mu\nu}w^\mu w^\nu-
30\left((\xi-{1 \over 15})R+m^2\right)w^2\right]~,
\end{equation}
\begin{equation}\label{2.53a}
h_d={r_d \over 1440\pi^2}\left[7w^2(w^2-2\nabla w)
-18(\nabla w)^2+22R_{\mu\nu}w^\mu w^\nu
-5(R+6m^2)w^2 \right]~~~.
\end{equation}
In these expressions $w^2=w^\mu w_\mu$, $\nabla w=\nabla^\mu w_\mu$, 
$\beta_l(x)=|g_{00}|^{1/2}\beta$
is the inverse local Tolman temperature, 
$a_{i,2}(x)$
are the second heat coefficients
of the 4-dimensional operators $L_i$, Eq.(\ref{3.1.7}).
The trace 
over spinor indexes in $a_{d,2}(x)$ 
is assumed.

The quantity $\Delta W_i$ which enters (\ref{2.51}) is
$$
\Delta W_i={\beta \over 16 \pi^{5/2}}\sum_{n=3}^{\infty}
c_{i,n}\Gamma\left(n-\frac 32\right)\zeta_R(2n-3)~
\bar{a}_{i,n}\left({\beta \over 2\pi}\right)^{2n-4}~~~.
$$
Here $c_{s,n}=1$, $c_{d,n}=1-2^{2n-3}$, $\zeta_R(z)$ is the Riemann
$\zeta$-function \cite{GrRy:94}. $\bar{a}_{i,n}$ are the coefficients
of asymptotic expansions of the heat kernels of the operators
$\bar{H}_i^2$. The quantity $\Delta W_i$ can also be rewritten in terms
of the physical metric and local temperature 
\cite{DoSc:88},\cite{DoSc:89}.  Let us emphasize that
except the term $W_i^{(3)}$ the high temperature 
expansion (\ref{2.51}) of the effective action has a local form. 
The non-local
contributions to the finite-temperature effective action was studied 
in \cite{GuZe:97}.

In the volume cutoff method the integration in
(\ref{2.51}) is performed over the spacetime of the
black hole
till the proper 
distance $\epsilon$
from the horizon. The action $W_i[g,\beta,\epsilon]$ and
free energy $F^E_i[g,\beta,\epsilon]$ diverge when
$\epsilon\rightarrow 0$. 
The divergences are caused only by the  
terms proportional to $\beta^{-4}$ and $\beta^{-2}$ in the functions 
$b_i(x,\beta)$.
The functions $h_i$ contribute 
to the divergence of the vacuum energy. 
As expected, 
the divergences in Euclidean and canonical free
energies coincide
\begin{equation}\label{2.100}
F^C_{i,\tiny\mbox{div}}[g,\beta,\epsilon]=
F^E_{i,\tiny\mbox{div}}[g,\beta,\epsilon]~~~.
\end{equation}
Here the quantities $F^C_{i,\tiny\mbox{div}}[g,\beta,\epsilon]$ 
are determined 
by 
Eqs. (\ref{2.5.1}),(\ref{2.5.2}) and 
$F^E_{i,\tiny\mbox{div}}[g,\beta,\epsilon]$ can be found
from the actions (\ref{2.51}) by subtracting
the vacuum energy, see Eq. (\ref{3.1.7a}).

\subsection{Conical
singularities and ultraviolet divergences}

We now consider the definition of the covariant
Euclidean effective
action on a complete manifold ${\cal M}_\beta$, i.e., 
on a manifold  
with conical singularities\footnote{There are several explicit 
calculations of an 
effective action
on some two-dimensional
\cite{FSW:96},\cite{FFZ:96a},\cite{FFZ:96b},
three-dimensional \cite{MaSo:97} and four-dimensional
\cite{FM:94},\cite{DFM:97b} spaces with conical singularities.
It is worth mentioning that conical singularities
appear in a large number of other physical applications
including cosmic strings \cite{Vilenkin:85},
orbifolds in string theory \cite{GSW:87}
and topological defects in superfluid phases of 
Helium 3 \cite{Volovik:97}.}.
The important feature of this case is that
the conical singularities result in additional
ultraviolet divergences depending on $\beta$.

It is convenient to introduce the wave operators
$\triangle_i=-\nabla^\mu\nabla_\mu + X_i$,
acting on the scalars and spinors. Here
$\nabla$'s are corresponding covariant derivatives.
We have
\begin{equation}\label{3.5.1}
L_s=\triangle_s+m^2~~~,
~~~L_d^2=\triangle_d+m^2~~~,
\end{equation}
where $X_s=(1/6-\xi) R$ and $X_d=\frac 14 R I$.
The analysis of Laplacian operators on cones and their heat 
kernels
can be found in Refs. \cite{Cheeger:83}-\cite{KaSt:91}. 
For physical applications the suitable 
representation
for the Green functions and heat kernels of
integer and half-odd-integer spins was given
by Dowker \cite{Dowker:77},\cite{Dowker:87},
see also Ref. \cite{DeJa:88}. These results are based on
the generalization of 
the representation obtained by Sommerfeld  
100 years ago \cite{Sommerfeld:1897}.

In general, a one-loop effective action can be defined 
by the Schwinger-DeWitt representation
\begin{equation}\label{3.5.2}
W[g,\beta]={\eta \over 2}\log\det(\triangle+m^2)=
-{\eta \over 2}
\int_{\delta^2}^{\infty}{ds \over s}
e^{-m^2s}\mbox{Tr}~e^{-s\triangle}~~~~,
\end{equation}
where $\delta^2$ is an ultraviolet cutoff. (For
briefness we omit the index $i$ indicating what type of the 
fields we are considering.) 
The geometrical structure
of the ultraviolet divergences is determined by the
first terms in the asymptotic expansion 
of the heat kernel at small $s$. 
On manifolds
without boundaries it takes the form
\begin{equation}\label{3.5.3}
\mbox{Tr}~e^{-s\triangle}\approx {1 \over (4\pi s)^{D/2}}
\left(B_0+sB_1+s^2 B_2+...\right)~~~,
\end{equation}
where $D$ is the dimension of ${\cal M}_\beta$.
The heat (Hadamard-Minackshisundaram-DeWitt-Seeley)
coefficients 
$B_k$ for $k\geq 1$ can be represented 
as the sum of two terms
\begin{equation}\label{3.5.4}
B_k=A_k+A_{\beta,k}~~~.
\end{equation}
Here $A_k$ has the form of the standard coefficient
defined on the regular domain of ${\cal M}_\beta$.
The term $A_{\beta,k}$ is an addition due to
conical singularities. This contribution is a functional
on $\Sigma$ which depends on the
geometrical characteristics of ${\cal M}_\beta$
near this surface. The first two coefficients $A_k$
and $A_{\beta,k}$
have the form
\begin{equation}\label{3.5.5}
A_1=d_1\int_{{\cal M}_\beta-\Sigma}R~~~,~~~
A_{\beta,1}=
{\pi \over 3\gamma}f_1(\gamma^2-1)
{\cal A}~~~,
\end{equation}
\begin{equation}\label{3.5.6}
A_2={1 \over 180}\int_{{\cal M}_\beta-\Sigma}
\left(q_1 R^{\mu\nu\lambda\rho}R_{\mu\nu\lambda\rho}
+q_2 R^{\mu\nu}R_{\mu\nu}+q_3 R^2\right)~~~,
\end{equation}
\begin{equation}\label{3.5.7}
A_{\beta,2}={\pi \over 3\gamma}\int_{\Sigma}
\left[(\gamma^4-1)p_1 {\cal P}+(\gamma^2-1)(p_2 R+p_3 {\cal Q})
\right]~~~,
\end{equation}
where $\gamma={\beta_H \over \beta}$. The coefficients 
$d_1$, $f_1$, $q_i$ and $p_i$
for spins $s=0$ and $1/2$ can be found in Table 1.

For a scalar field the coefficient $A_{\beta,1}$  
was first found  by Cheeger \cite{Cheeger:83}, 
see also Refs. \cite{CKV:94},\cite{Fursaev:94a}.
The coefficients $A_{\beta,1}$  for spins $1/2$ and $1$ were obtained
in Refs. \cite{Kabat:95},\cite{FM:97} and for
spins $3/2$ and $2$ in \cite{FM:97}. 
The coefficient $A_{\beta,2}$ 
and general structure
of the higher coefficients $A_{\beta,k}$
were analysed in Refs. \cite{Fursaev:94b}, 
\cite{Dowker:94a} and \cite{Dowker:94b} for
scalars, in Ref. \cite{Fursaev:97}, and for spinors
in Ref. \cite{DFM:97a}.

\bigskip

The divergent part $W_{\tiny\mbox{div}}[g,\beta]$
of the effective action
on ${\cal M}_\beta$ can be written in different regularizations. It is 
convenient to begin with the dimensional regularization.
For $D\neq 4$ one finds
from Eqs. (\ref{3.5.2}) and (\ref{3.5.3})
\begin{equation}\label{3.5.8}
W_{\tiny\mbox{div}}[g,\beta,D]=
-{\eta \over 2}
\int_{0}^{\infty}{ds \over s}
e^{-m^2s}{1 \over (4\pi s)^{D/2}}
\left(B_0+s B_1+s^2 B_2\right)~~~.
\end{equation}
The Euclidean free energy is obtained from the
action by subtracting the vacuum energy (see 
Eq. (\ref{3.1.7a})). On a regular Euclidean manifold the
divergences are determined only
by the coefficients $A_0$, $A_1$, $A_2$ and they are proportional 
to the period $\beta$. So they do not contribute 
to the free energy and the entropy. In case of conical singularities
the divergences have the form of polynomials in powers
of  $\beta^{-1}$ because of the additional terms 
$A_{\beta,k}$. These terms are not removed by subtracting
the vacuum part. For this reason the Euclidean
free energy is divergent.
The divergent part $F^E_{\tiny\mbox{div}}[g,\beta,D]$ can
be easily found with the help of Eqs. 
(\ref{3.5.5}), (\ref{3.5.7}) and (\ref{3.5.8}).

As follows from  Eq. (\ref{2.6.4}),   the ultraviolet divergences of
the canonical and Euclidean free energies for scalar and spinor  fields
are {\it identical} in the dimensional regularization. This
coincidence  also takes place  in the Pauli-Villars regularization, so
one can write 
\begin{equation}\label{3.5.8a}
F^E_{i,\tiny\mbox{div}}[g,\beta,\delta]=
F^C_{i,\tiny\mbox{div}}[g,\beta,\delta]~~~,~~~i=s,d~~~.
\end{equation}
where $\delta$ is a regularization parameter 
($\delta=D-4$ for the dimensional regularization
and $\delta=\mu^{-1}$ for PV regularization).
As the important consequence of (\ref{3.5.8a}),
the divergences of the entropy in 
different formulations coincide as well
\begin{equation}\label{3.5.8b}
S^E_{i,\tiny\mbox{div}}[g,\beta,\delta]=
S^C_{i,\tiny\mbox{div}}[g,\beta,\delta]~~~,~~~i=s,d~~~.
\end{equation}

\subsection{Relation between canonical and Euclidean
formulations in the presence of horizons}

Till now we discussed and compared the divergent parts of 
the free energies calculated in the canonical and covariant
Euclidean formulations in the presence of a horizon.
In this Section we make remarks concerning the relation 
between the finite parts of the free energies in these
formulations.

As we discussed earlier, the canonical and Euclidean
formulations are completely 
equivalent for static spacetimes 
without horizons. This conclusion is based 
on the fact that 
the effective actions $W[g,\beta]$
and $\bar{W}[g,\beta]$ given on
${\cal M}_\beta$ and $\bar{\cal M}_\beta$, respectively,
are related by a conformal transformation. 
The Euclidean and canonical
free energies obtained from $W[g,\beta]$
and $\bar{W}[g,\beta]$ by subtracting zero
temperature parts do not have the divergences.
They are free from the anomalies
caused by renormalization and as the result
$F^E$ and $F^C$ coincide in the absence of a horizon.

The total functionals $F^C[g,\beta,\epsilon]$ and
$F^E[g,\beta,\epsilon]$ (containing both $\epsilon$-divergent 
and regular parts) also coincide
on the backgrounds with horizons provided one uses
the volume cutoff method. In this method the conformal 
transformation which relates the free energies is well-defined.

The difficulties arise when one compares
two formulations on a complete background which
includes the horizon. 
In this case the spaces
${\cal M}_\beta$ and $\bar{{\cal M}}_\beta$,
have different topologies, $\R^2\times \Sigma$ and
$S^1\times \bar{\cal B}$, respectively,
and the conformal 
transformation of 
${\cal M}_\beta$ onto $\bar{{\cal M}}_\beta$ is 
singular on the 
bifurcation surface. For this reason the relation
between the two formulations requires an additional
analysis.

We showed that there exist ultraviolet regularizations which are
applicable to both covariant Euclidean 
and canonical free energies. Let us consider for example 
a situation when volume cutoff and Pauli-Villars
regularizations are applied simultaneously.
Then the free energies depend on $\epsilon$ and 
PV parameter $\mu$. Because the horizon is
excluded  there is  the equality
\begin{equation}\label{6.3a}
F_{i}^E[g,\beta,\mu,\epsilon]=
F_{i}^C[g,\beta,\mu,\epsilon]~~~.
\end{equation}
As we have shown earlier, the left and right parts of this
equality remain finite when $\epsilon$ cutoff is removed.
From (\ref{6.3a}) we obtain 
\begin{equation}\label{6.3}
F_{i}^E[g,\beta,\mu]=\lim_{\epsilon\rightarrow 0}
F_{i}^E[g,\beta,\mu,\epsilon]=
\lim_{\epsilon\rightarrow 0}
F_{i}^C[g,\beta,\mu,\epsilon]=F_{i}^C[g,\beta,\mu]~~~.
\end{equation}
In the limit $\epsilon\rightarrow 0$ $F_{i}^E$ becomes the functional 
defined on ${\cal M}_\beta$, whereas in the same limit
$F_{i}^C$  is defined on 
$\bar{{\cal M}}_\beta$. Equation (\ref{6.3}) supports
the conclusion \cite{Solod:96a} and \cite{Fursaev:97}
that not only the divergencies, but also
the complete {\it bare} free energies in the covariant
Euclidean and canonical formulations 
are {\it equivalent} when 
one uses the ultraviolet regularization.

There are examples where 
equality (\ref{6.3}) can be checked 
by explicit calculations. 
For instance, by rewriting the density of levels 
(\ref{2.57}) of massive 
scalar fields in the Rindler space-time 
in the Pauli-Villars
regularization one can find from it the canonical free energy and
confirm equality (\ref{6.3}). An explicit derivation
of Eq. (\ref{6.3}) is also possible in two dimensions
\cite{Solod:96a}.

Relation (\ref{6.3}) demonstrates that the partition function 
$Z^E$ defined by 
Euclidean path integral (\ref{3.1.1}) 
coincides\footnote{More precisely, to equate $Z^C$ 
and $Z^E$
one has to normalize $Z^E$ so that  to
exclude the vacuum energy.}
with the canonical partition function $Z^C$, Eq. (\ref{2.24}),
including the case of spacetimes with horizons.
This fact is very important because it justifies the 
statistical-mechanical 
interpretation of the Gibbons-Hawking path integral method 
\cite{GiHa:76}, \cite{Hawk:79} when it is applied
to black holes.

It should be noted that (\ref{6.3})  
was analysed for scalar and spinor fields only \cite{Fursaev:97}. 
The relation between the canonical and the
covariant Euclidean 
formulations for high spin fields needs an additional 
consideration. 
Some discussion of vector fields can be found 
in Refs. \cite{MoIe:97}-\cite{Moretti:97}.

\subsection{Remarks on rotating and extremal black holes}
\setcounter{equation}0

So far our discussion was restricted by  static 
black hole geometries. The covariant Euclidean formulation of 
statistical mechanics 
for rotating black holes
was studied in Refs. \cite{MaSo:96} and \cite{MaSo:97}. 
It was shown in \cite{MaSo:96}, an Euclidean manifold 
which is
obtained by the Wick rotation of a stationary geometry
with the Killing horizon
has a conical singularity similar to the one which 
appears in static spaces. This singularity results in 
the one-loop divergence which
has the form (\ref{3.5}). 
The canonical formulation of the statistical mechanics of quantum fields
on a rotating black hole background
was discussed in 
Refs. \cite{IcSa:95}-\cite{HKP:97}.
However, the relation between the canonical and 
covariant
Euclidean formulations in this case has not been
investigated.

\bigskip

Some remarks are also in order about extremal black holes.
The horizon of an extremal black hole has  
zero surface gravity which indicates that the temperature 
of such a black hole
is zero  \cite{AHL:95},\cite{Vanzo:97}.
In the Euclidean theory extremal black holes have the topology of an 
annulus and there is no conical
singularity for an arbitrary period $\beta$ 
\cite{HHR:95},\cite{Teitelboim:95}. 
The statistical mechanics of quantum fields 
on the background of the extremal black hole
has important features which differs it from the non-extremal case.
In particular, 
the leading divergence of the
density of levels ${d n \over d\omega}$ 
of quantum fields is proportional to $e^{L/M}$ where
$M$ is the black hole mass and $L$ is the 
proper distance between the 
spatial boundary and the location of the 
cutoff near the horizon \cite{GhMi:95},\cite{CVZ:95b}.
Nevertheless,  the ultraviolet type regularizations 
still can be applied 
to eliminate the divergencies of the density of states 
and canonical free energy \cite{DLM:95}. Interestingly, the divergences of
${d n \over d\omega}$ and $F^C$ in these
regularizations have the form which is similar
to Eqs. (\ref{2.6.1})-(\ref{2.6.4})
and (\ref{33.1}),(\ref{33.2}),(\ref{3.5}).
There are no results comparing the canonical and 
covariant Euclidean 
formulations
of statistical mechanics for extremal black holes.

\section{Thermodynamics and statistical mechanics
of black holes}
\subsection{Statistical-mechanical entropy and quantum correction
to black hole entropy}

\setcounter{equation}0

Till now we analysed the properties of the  statistical-mechanical
entropy  of quantum fields around a black hole.  In the general case
this entropy is divergent while the Bekenstein-Hawking entropy is
finite.  What is the relation between the statistical mechanical
entropy of quantum excitations of a black hole and its thermodynamical
entropy?  

In quantum field theory the quantum corrections  are ultraviolet
divergent quantities whose  divergences are removed by the
renormalization of bare coupling constants.  As we demonstrated the
divergencies of the statistical-mechanical entropy
$S^C_{\tiny\mbox{div}}$ have ultraviolet form. It was suggested in
\cite{SuUg:94}  that $S^C_{\tiny\mbox{div}}$ can be absorbed by the
standard  renormalition of the Newton constant.

Let us discuss this renormalization approach in more details. The
complete information concerning the canonical ensemble of black holes
with a given inverse temperature $\beta$ at the boundary is contained
in the partition function $Z(\beta)$ given by the Euclidean path
integral \cite{Hawk:79}
\begin{equation}\label{2.2x}
Z(\beta)=\int [D\Phi] \exp (-I [\Phi ]) .
\end{equation}
Here the integration is taken over all fields including the
gravitational one that are given on the Euclidean section and are
periodic (or antiperiodic) in the imaginary time coordinate $\tau$ with 
period
$\beta$. The quantity $\Phi$ is understood as the collective
variable describing  the fields. In particular, it contains the
gravitational field. Here $[D\Phi]$ is the measure of the space
of fields $\Phi$ and $I$ is the Euclidean action of the field
configuration.  The action $I$ includes the Euclidean Einstein
action. The state of the system is determined by the choice of the
boundary conditions on the metrics that one integrates over. For
the canonical ensemble of the gravitational field inside a
spherical box of radius $r_0$ at temperature $T$ one must integrate
over all the metrics inside $r_0$ which are periodically identified
in the imaginary time direction with period $\beta=T^{-1}$. Such a
partition function must describe in particular the thermal ensemble
of black holes.  The partition function $Z$ is related to the
effective action $\Gamma=-\ln Z$. The free energy $F$
is defined as 
$F=\beta^{-1}\Gamma=-\beta^{-1}\ln Z$.

By using the stationary-phase approximation one gets
\begin{equation}\label{2.3x}
\beta F\equiv \Gamma = I [\Phi_0] +W +\ldots  .
\end{equation}
Here $\Phi_0$ is the (generally speaking, complex)  solution of
classical field equations for action $I[\Phi]$ obeying the required
periodicity and boundary conditions. Besides the tree-level
contribution $I[\Phi_0]$, expression (\ref{2.3x}) includes
also one-loop corrections  $W$, connected with the
contributions of the fields perturbations on the background
$\Phi_0$, and higher order terms in loop expansion, denoted
by $(\ldots )$. For free fields, $W$ is a one-loop
effective action computed in the covariant Euclidean formulation
of the theory, see Eqs. (\ref{3.1.6}).

The one-loop divergences 
appearing in $W$ can be absorbed by
the renormalization of the couplings of the initial classical
action $I$. To this aim the latter is chosen in the 
form
\begin{equation} \label{b}
I(G_B,\Lambda_B,c^i_B)=\int d^4x \sqrt{g}L ,
\end{equation}
\begin{equation}\label{bb}
L=\left[ -\frac{\Lambda_B}{8\pi G_B} -\frac{R}{16\pi G_B}
+c^1_B R^2 +c^2_B R_{\mu\nu}R^{\mu\nu} +c_B^3 
R_{\alpha\beta\mu\nu}R^{\alpha\beta\mu\nu}
\right]~~~.
\end{equation}
 
In the presence of one-loop divergences the stationary-phase
approximation procedure is modified as follows. Denote by
$W_{\tiny\mbox{div}}$ the UV-divergent part of the one-loop effective 
action $W$. Then the renormalized quantities are 
defined as 
\begin{equation}
I_{\tiny\mbox{ren}}\equiv I(G_{\tiny\mbox{ren}},\Lambda_{\tiny\mbox{ren}},
c^i_{\tiny\mbox{ren}})=I(G_B,\Lambda_B,c^i_B)+
W_{\tiny\mbox{div}}~~~,\hspace{0.5cm}
W_{\tiny\mbox{ren}}=W-W_{\tiny\mbox{div}}~~~.
\end{equation}
Now the starting point of the semiclassical approximation is in finding
the extremal of the renormalized
action $I_{\tiny\mbox{ren}}$. Since for this background
$W_{\tiny\mbox{ren}}$ is finite and proportional to $\hbar$, this part of
the action describes small quantum corrections. 

The key observation of the renormalization approach is that
$W_{\tiny\mbox{div}}$ has the same structure as (\ref{b})--(\ref{bb}) and
hence $W_{\tiny\mbox{div}}$ can be absorbed   by 
simple
redefinition of the coupling constants in 
$I(G_B,\Lambda_B,c^i_B)$. In other words, 
$I_{\tiny\mbox{ren}}$ is identical to the initial classical action
$I$ with the only change that the bare coefficients
$\Lambda_B$, $G_B$, and $c^i_B$ are substituted by their renormalized
versions $\Lambda_{\tiny\mbox{ren}}$, $G_{\tiny\mbox{ren}}$, and
$c^i_{\tiny\mbox{ren}}$.
$W_{\tiny\mbox{div}}$ can be found by using relation 
(\ref{3.5.3}) and expressions (\ref{3.5.4})-(\ref{3.5.6}) for the 
heat kernel coefficients of the corresponding Laplace operators 
(on regular backgrounds).
The relation between bare and renormalized couplings depends 
on the regularization.
For instance, in PV regularization 
the 
renormalization of the Newton constant for the non-minimally
coupled scalar field results is   
\begin{equation}\label{x1}
{1\over G_{\tiny\mbox{ren}}}={1\over G_{B}}+{c \over 2\pi}
\left(\frac 16-\xi\right)\mu^2~~~
 ,
\end{equation}
where $c=\ln{729 \over 256}$ and $\mu$ is the PV cutoff.

The free energy $F(\beta)=\beta^{-1}\Gamma$ being expressed in terms of
the renormalized constants is  finite. By calculating $F(\beta)$ on the
black hole instanton one finds the one-loop free energy of a black
hole. The ``observable'' thermodynamic entropy of a black hole $S^{TD}$
\cite{Frolov:95}  has the standard form $S^{TD}=\beta^2d F(\beta)/d
\beta$. If we neglect for a while  by the logarithmic divergencies and
put $\Lambda_{\tiny\mbox{ren}}=0$ we obtain
\begin{equation}\label{3.7.6aa}
S^{TD}=S^{BH}(G_{\tiny\mbox{ren}})+O(\hbar)~~~.
\end{equation}
Here $S^{BH}(G_{\tiny\mbox{ren}})$ is the Bekenstein-Hawking 
entropy in the theory of general
relativity
with the Newton constant $G_{\tiny\mbox{ren}}$. The terms 
$O(\hbar)$ represent finite quantum
corrections proportional to $\hbar$. Equations 
(\ref{x1}) and (\ref{3.7.6aa}) can be compared
with the statistical mechanical entropy $S^C$ of quantum
fields. For a scalar field the leading divergency
of $S^C$ can be found from (\ref{3.6b})
\begin{equation}\label{x2}
S^C_{\tiny\mbox{div}}={c \over 48\pi}\mu^2 {\cal A}~~~.
\end{equation}
Thus, by using Eq. (\ref{x1}) we find
\begin{equation}\label{3.7.6ab}
S^{BH}(G_{\tiny\mbox{ren}})=
S^{BH}(G_{\tiny\mbox{B}})+S^C_{\tiny\mbox{div}}-
Q_{\tiny\mbox{div}}~~~,
\end{equation}
where the quantity $Q_{\tiny\mbox{div}}=\xi c\mu^2 {\cal A}/(2\pi)$
appears when a scalar field is non-minimal coupled with the
curvature. In the next Section we show that Eq. (\ref{3.7.6ab}) 
preserves its form when not only
the leading but all the divergencies are included.

\bigskip

Equation (\ref{3.7.6aa}) explicitly demonstrates
that the ``observable'' Bekenstein-Hawking entropy contains the
statistical-mechanical entropy of black-hole's quantum excitations as
its part, but in the general case it does not coincides with it. For
non-minimal coupling an additional term $Q$ is present. Even in the
absence on nonminimal coupling when $Q=0$, the
presence of the bare pure geometrical contribution
$S^{BH}(G_{\tiny\mbox{B}})$ evidently excludes the
possibility to identify
$S^{BH}(G_{\tiny\mbox{ren}})$ with
$S^{C}_{\tiny\mbox{div}}$ which has clear statistical mechanical
meaning. Moreover, in order to have finite value of
$S^{BH}(G_{\tiny\mbox{ren}})$ one needs to assume
that pure geometrical ``entropy''
$S^{BH}(G_{\tiny\mbox{B}})$ is infinite and
negative. For this reason the idea to relate $S^{BH}$ with quantum
excitations does not work, at least in the standard renormalization
approach. The way out of this problem is to restrict oneself by
considering special class of the theories where
$S^{BH}(G_{\tiny\mbox{B}})=0$. It happens when
$G^{-1}_{\tiny\mbox{B}}=0$, and hence initially gravity
is not dynamical. The dynamics of the gravitational field arises as the
result of quantum effects. The induced gravity is an example of
such a theory.

\section{General renormalization and the Noether
charge}
\setcounter{equation}0

Before discussing the models of induced gravity
we make comments on the generalization of Eq. (\ref{3.7.6ab}).
Let us consider the case 
when terms quadratic in curvature are preserved in the 
renormalized action. In this case the classical black hole
entropy  is
(see Refs. \cite{FS:95}--\cite{JKM:94})
\begin{equation}\label{3.6.4}
S^{BH}(G_{\tiny\mbox{ren}},c^i_{\tiny\mbox{ren}})=
{1 \over 4 G_{\tiny\mbox{ren}}}
{\cal A}-
\int_{\Sigma}\sqrt{\sigma}
d^2\theta~
(8\pi c^1_{\tiny\mbox{ren}} R+4\pi c^2_{\tiny\mbox{ren}} {\cal Q}+ 
8\pi c^3_{\tiny\mbox{ren}} {\cal R})~~~.
\end{equation}
The integral in (\ref{3.6.4}) is taken over
the bifurcation surface of the horizon. The first term in the r.h.s. of
Eq. (\ref{3.6.4}) is the Bekenstein-Hawking entropy, and other terms
are additions because of the high curvature terms in the action. 

For PV and dimensional regularizations in the model with the 
scalar and spinor fields
the relation (\ref{3.7.6ab}) takes the 
form\footnote{
In Section 4.3 we pointed out that the divergencies 
$S^C_{\tiny\mbox{div}}$ of the statistical-mechanical  entropy in the
volume cutoff method also take the ultraviolet  form if one identifies
the
volume cutoff parameter $\epsilon$ with an ultraviolet cutoff.  The
problem is that  each field species requires  its particular relation
between $\epsilon$ and $\mu$, and there is no universal relation which 
enables one to remove the divergencies from all fields. For instance,
for scalars the relation between the volume cutoff and PV parameters 
looks as $\epsilon^{-2}=c{15 \over 2}\mu^2$, see Eqs. (\ref{2.5.4a}),
(\ref{3.6b}), while for spinors the same relation has to be 
$\epsilon^{-2}=c{15 \over 11}\mu^2$. This indicates that the
volume cutoff parameter is not a truly covariant 
ultraviolet regulator. 
Thus, regarding the  renormalization problem, the volume cutoff and
``brick-wall'' methods run into the difficulty. Some other difficulties
of these methods are discussed in Refs.
\cite{Emparan:95}--\cite{BeMa:96}. For two-dimensional black hole models
the relation between the ``brick wall'' results and the renormalized
quantum correction to black hole entropy is discussed in  Refs.
\cite{FFZ:96a} and \cite{FFZ:96b}.}
\begin{equation}\label{3.7.6b}
S^{BH}(G_{\tiny\mbox{ren}},c^i_{\tiny\mbox{ren}})=
S^{BH}(G_{\tiny\mbox{B}},c^i_{\tiny\mbox{B}})
+S^{C}_{\tiny\mbox{div}}-{Q}_{\tiny\mbox{div}}~~~.
\end{equation}
In PV regularization $S^C_{\tiny\mbox{div}}$ is determined
by expression (\ref{3.6b}). The quantity 
${Q}_{\tiny\mbox{div}}$ appears for nonminimally
coupled scalars,
\begin{equation}\label{divq}
{Q}_{\tiny\mbox{div}}=\xi {1 \over 8\pi}\int_{\Sigma}
\left[b+a\left(\frac 16-\xi\right)R 
\right]~~~.
\end{equation}
The coefficients $a$, $b$ depend on PV cutoff $\mu$ and are given by Eqs.
(\ref{3.5b}), (\ref{3.5c}).
When ${Q}_{\tiny\mbox{div}}=0$ the general
proof of  (\ref{3.7.6b}) was given in Refs.  \cite{DLM:95} and 
\cite{FS:96} (see also Refs. \cite{LaWi:96}, \cite{Solod:95b}, 
\cite{Fursaev:95} and \cite{Kabat:95}). 
 
For scalar fields with the nonminimal
coupling, quantity (\ref{divq})
can be written as
\begin{equation}\label{x3}
{Q}_{\tiny\mbox{div}}=2\pi\xi \int_{\Sigma}
\langle \hat{\phi}^2 \rangle_{\tiny\mbox{div}}~~~,
\end{equation}
where it is assumed that the fluctuation of the
scalar field
$\langle \hat{\phi}^2 \rangle$ is computed in PV regularization. 
Relation (\ref{x3})
was first found by Solodukhin \cite{Solodukhin:95}
with the help of the Euclidean formulation 
of the theory with conical singularities.

The reason why quantity (\ref{x3}) appears in 
formula (\ref{3.7.6b}) is the following. In the presence
of scalar fields with nonminimal couplings the 
Bekenstein-Hawking entropy includes the additional term
$Q=2\pi \xi \int_{\Sigma} \phi^2$, where $\phi$ is the
classical field, see \cite{JKM:94}. In quantum theory the 
quantity $Q$ becomes the operator
whose average has the divergent part which coincides 
with ${Q}_{\tiny\mbox{div}}$.
In the one-loop approximation 
${Q}_{\tiny\mbox{div}}$ determines the quantum correction to
$S^{BH}$ because of the non-minimal coupling.

\bigskip

It was shown recently by Wald {\it et al.} \cite{Wald:93}--\cite{JKM:94}, 
\cite{Nelson:94}, \cite{Brown:95}
that the
classical black hole entropy in diffeomorfism invariant
theories of gravity can be interpreted as the
Noether charge. In Appendix we demonstrate that $Q$ 
coincides with the Noether charge for the nonminimally
coupled matter fields propagating on the fixed curved background and 
construct the corresponding
Noether current. We also show that
\begin{equation}\label{3.8.10}
Q={2\pi \over \kappa}(H-E)~~~,
\end{equation}
where $\kappa$ is the surface gravity, $H$ is the canonical
Hamiltonian of the fields and $E$ is the energy of the
fields obtained from the stress-energy tenor,
see definitions (\ref{3.8.5}) and (\ref{3.8.6a}) of Appendix.
The latter relation plays an important role in the
models of induced gravity.

\section{Black hole entropy in induced gravity}
\setcounter{equation}0

The theory of induced gravity was suggested by Sakharov \cite{Sakh:68}
long ago. The low-energy gravitational effective action $\Gamma[g]$ in
this theory is defined as a quantum average of the constituent fields
$\Phi$ propagating in a given external  gravitational background $g$
\begin{equation}\label{2}
\exp(-\Gamma[g])=\int [D \Phi] \exp(-I[g,\Phi])~~~.
\end{equation}
The Sakharov's basic assumption is that the  gravity becomes dynamical
only as the result of quantum  effects of the constituent  fields.
As a result, we have theories of the special type, namely theories
with $I[g]=0$. 
 The
gravitons in such a picture are analogous to the phonon field
describing collective excitations of a crystal lattice in the
low-temperature limit of the theory. In the general case,
each particular constituent field 
in (\ref{2}) gives a divergent contribution to the effective action
$\Gamma[g]$. In the one loop approximation the divergent terms are
local and of the zero order, linear and quadratic in curvature. In the
induced gravity the constituents obey additional constraints, so that
the divergences cancel each other. It is also assumed that some of the
fields  have masses comparable to the Planck mass and the constraints
are chosen so that the induced cosmological constant vanishes. As the
result the effective action $\Gamma[g]$ is finite and in the low-energy
limit it has the form of  the Einstein-Hilbert action
\begin{equation}\label{3}
\Gamma[g]=-{1\over 16\pi G}\left(\int_{\cal M} dV\, 
\, R +2\int_{\partial  {\cal
M}}dv \, K\right)+
\ldots \, \,
,
\end{equation}
where Newton's constant $G$ is determined by the masses of the heavy
constituents\footnote{As an  another realization of the idea of induced
gravity it is worth mentioning the approach by Adler,  for a review  see
Refs. \cite{Adler:82},\cite{NoVa:91}. According to Adler, the induced
Newton constant  may arise as a result of the dimensional transmutation
because of the dynamical symmetry breaking in the underlying  scale
invariant theory. This picture is different  from the mechanism under
discussion.}. The dots  in (\ref{3}) indicate possible higher 
curvature corrections to $W[g]$ which are suppressed  by the
power factors of  $m_i^{-2}$ when the curvature is small. The vacuum
Einstein  equations $\delta W / \delta g^{\mu\nu}=0$ are equivalent to 
the requirement that the vacuum expectation values of the total
stress-energy  of the constituents vanishes
\begin{equation}\label{4}
\langle \hat{T}_{\mu\nu}\rangle
=  0\, .
\end{equation}
The value of the Einstein-Hilbert action (\ref{3}) calculated on the
Gibbons-Hawking  instanton determines the classical free energy of the
black hole, and hence  gives the Bekenstein-Hawking entropy $S^{BH}$. 

Consider a model of induced gravity  \cite{FFZ:97} that consists of a
number of scalar fields with masses $m_s$ and a number of Dirac
fermions with masses $m_d$. Scalar fields can have non-minimal
couplings $\xi_s$.  It is convenient to introduce the following two
functions
\begin{equation}\label{fro-17}
p(z)=\sum_s m_s^{2z} -4\sum_d m_d^{2z}\, ,\hspace{.5cm}
q(z)=\sum_s m_s^{2z}(1-6\xi_s) +2\sum_d m_d^{2z}~~~.
\end{equation}
Direct
calculations show that the induced cosmological  constant vanishes when
\begin{equation}\label{fro-18}
p(0)=p(1)=p(2)=p'(2)=0~~~.
\end{equation}
The induced gravitational coupling constant $G$ is finite if  the
following constraints are satisfied
\begin{equation}\label{fro-18a}
q(0)=q(1)=0~~~.
\end{equation}
Relations (\ref{fro-18}) are satisfied for 
theories with supersymmetric massive multiplets for which $p(z)=0$.
Equations  (\ref{fro-18a}) form a linear system
defining $\xi_s$.

The presence of the non-minimally coupled constituents
is important. In this case it is possible to satisfy
constraints (\ref{fro-18a}) on the parameters $m_s$, $m_d$
and $\xi_s$ which guarantee
the cancellation of the leading ultraviolet
divergencies of the induced gravitational action
$\Gamma[g]$.
The induced Newton's constant in this model 
is \cite{FFZ:97}
\begin{equation}\label{7}
{1 \over G}=
{1 \over 12\pi} \left(\sum_{s}(1-6\xi_s)~ m_s^2 
\ln m_s^2+2\sum_{d}
m^2_d\ln m_d^2\right)~~~.
\end{equation}

Let us analyze now the entropy of a Schwarzschild black hole in the
induced gravity. If constraints (\ref{fro-18}) and (\ref{fro-18a}) are
satisfied the induced action has logarithmically divergent terms only.
However these terms are quadratic in curvature and on the Schwarzschild
background they  can be neglected because they do not depend on the
geometry.  Consider the difference $S^C-Q$, where $S^C$ is the
statistical-mechanical  entropy of the scalar and spinor massive fields
(constituents) and $Q$ is the Noether charge which appears because of
non-minimal couplings of the scalar constituents. $S^C$ and $Q$ are
assumed to be regularized  according to the same prescription, say, by
PV method. In the low-energy limit of the theory one can  calculate
these quantities in the leading order in $m_i^2$. In this approximation
$S^C$  coincides with the part  $S^C_{\tiny\mbox{div}}$ determined in
PV-regularization by Eq. (\ref{3.6a}). Analogously, the Noether charge
is approximated by $Q_{\tiny\mbox{div}}$, which can be found from Eq.
(\ref{divq}). It is easy to check that for a Schwarzschild  black hole
the leading divergencies  in $S^C-Q$ are cancelled and from Eqs.
(\ref{3.6a}) and (\ref{divq}) one obtains
\begin{equation}\label{8c}
S^C-Q\simeq S^C_{\tiny\mbox{div}}-Q_{\tiny\mbox{div}}=
{1 \over 4G}{\cal A}+C~~~.
\end{equation}
Here $C$ is
a numerical divergent  constant\footnote{The divergence of $C$ is a
specific property of the given simplified model which is not free from
some logarithmic divergencies. In a more complicated model including,
for instance, vector massive fields $C$ can be made finite.}.  In our
consideration $C$ can be neglected because it does not depend on the
black hole geometry.  Thus, up to this numerical addition, $S^C-Q$ agrees
precisely with the induced Bekenstein-Hawking  entropy $S^{BH}={1 \over
4G}{\cal A}$. The result (\ref{8c}) can be also 
obtained by using the
dimensional  regularization\footnote{It should be noted, however,  that
the check of Eq. (\ref{8c}) in the volume cutoff regularization may
meet  a difficulty \cite{MoIe:97b}. Presumably, it happens because this
regularization makes the background space incomplete, and so using 
expression (\ref{x3}) for $Q$ in the entropy becomes 
unjustified.}. 

As follows from (\ref{8c}), the Bekenstein-Hawking  entropy is
determined by the leading order  of the
statistical-mechanical entropy of the heavy constituent fields. The
quantity $S^C_{\tiny\mbox{div}}$ comes out as the result
of integration over a narrow layer located
near the horizon and having the size of the order of the Comtpton wave 
length of
the constituents, which is comparable to  the Planck length. Thus, the black
hole entropy depends only on the local properties of the horizon. This
conclusion implies that relation (\ref{8c}) must be also valid  for a
generic static or stationary black holes. 

Another conclusion which can be drawn from (\ref{8c})
is that the black hole entropy $S^{BH}$ is not identical to
the statistical-mechanical entropy $S^C$. Actually, $S^{BH}$
is a finite quantity, while $S^C$ diverges 
because each of the constituents gives a positive divergent contribution 
into it. 
In Eq. (\ref{8c}) 
the divergences of $S^C$ are eliminated by subtraction
of the charge $Q$. As it can be shown, in the model 
in question $Q$ has positive divergencies.
Thus, the presence of the fields with non-minimal
couplings which is imperative in order to make the theory
ultraviolet finite also provides the subtraction
of the divergencies of $S^C$.

\bigskip

Is there a statistical-mechanical explanation
why the Noether charge $Q$ enters relation (\ref{8c})?
A possible mechanism
was suggested in Ref. \cite{FF:97a}.
The  induced gravity explains
the black hole entropy by relating it to 
statistical-mechanical entropy of the constituents. As the result
of the backreaction effect, the black hole
geometry responds to quantum
fluctuations of these fields. One consequence of this 
effect is that fluctuations of the energy
of fields $E$ near the average value $E=0$ cause
the fluctuations of the black hole mass 
near its average value $M$.
The black hole entropy $S^{BH}$ determines the number density of 
the states in the interval $(M,M+\Delta M)$ so 
the quantity $\exp S^{BH}$ 
coincides with the degeneracy $\nu(M)$ of the 
black hole mass 
spectrum.
Now it is easy to show by using differential mass 
formula \cite{BCH:73} (see also Eq. (11.2.48) of 
Ref. \cite{FN:book}) that in the
leading order in the Planck constant
the change $\Delta M$
of the black hole mass 
coincides with
the change $\Delta E$ of the energy of the constituents.
Therefore, the problem of finding the degeneracy
of the black hole mass spectrum can be reduced to the
problem of finding the spectrum of the energy of the
constituent 
fields, which is more simple. In particular, 
the black hole entropy can be related with the 
number 
density $\nu(E)$ of $E$ as $e^{S^{BH}}=\nu(E=0)$.

Contrary to $S^{BH}$, the statistical-mechanical entropy
$S^C$ is related to the spectrum of
the Hamiltonian $H$ generating translations
of the system along the Killing time.
$H$ differs from $E$ by the  Noether charge $Q$.
That is why $S^{BH}$ and $S^C$ are different and
related by Eq. (\ref{8c}) \cite{FF:97a}.

Why the density number of states with the energy $E$ is finite, while
the number of states of $H$ diverges even in the ultraviolet finite
theory?  The answer to this question is related to the specific
property of the quantum system in the  presence of the Killing horizon.
As we pointed out, the spectrum of frequencies  of the single-particle
excitations in this case does not have the mass gap. 
In other words, there
exist modes with negligibly small frequencies, so called {\it soft
modes} \cite{FF:97a}. 
An arbitrary number of the soft modes can be added to such a
system without changing its canonical energy $H$.  It is the soft modes
which result in additional infinite degeneracy of the spectrum of the
Hamiltonian $H$. However, contribution of the soft modes to the energy
$E$ is not zero and the spectrum of $E$ doesn't have the  redundant 
degeneracy.

The reason why the soft modes contribute to $E$ is that $E$ differs
from $H$ by the Noether charge  $Q$. In fact, 
$Q$ is determined by the soft modes  only. To see this, it is
sufficient to approximate the black hole geometry (\ref{rind}) near the
horizon by the Rindler space (\ref{rind}).  According to Eq.
(\ref{x3}), $Q$ is determined by the average of the scalar
correlator $\langle \hat{\phi}^2 \rangle$ on the bifurcation surface
$\Sigma$. In the Rindler approximation   the correlator can be easily
computed \cite{FF:97a}
$$
\langle \hat{\phi}(x)\hat{\phi}(x')\rangle
=\mbox{Tr}\left[\hat{\rho}
\hat{\phi}(x)\hat{\phi}(x')\right]
$$
\begin{equation}\label{soft1}
=\int_{0}^{\infty}d\omega\int d^2k\left[
n_\omega~ \phi^{*}_{\omega,k}(x)\phi_{\omega,k}(x')+
(n_\omega+1)\phi_{\omega,k}(x)\phi^{*}_{\omega,k}
(x')\right]~~~.
\end{equation}
Here $n_\omega=(\exp(2\pi\omega/\kappa)-1)^{-1}$ is the number of 
particles 
at the Hawking temperature with the
energy $\omega$. The
functions $\phi_{\omega,k}(x)$ are the eigen functions 
of the single-particle Hamiltonian $H_s$ determined by Eq. (\ref{2.3}).
On the Rindler space (\ref{rind})
\begin{equation}\label{soft2}
\phi_{\omega,k}(x)={1 \over 4\pi ^3 \kappa^{1/2}}
\left(\sinh (\pi \omega/\kappa)\right)^{1/2}~
K_{i\omega/\kappa}(\rho(m^2+k_j^2)^{1/2})e^{-i\omega t}
e^{-ik_jz^j}~~~,
\end{equation}
where $k^i$ and $z^i$ are the momentum and the coordinate 
along the horizon. $K_{i\omega/\kappa}$ are the
modified Bessel functions. When the proper distance to the horizon $\rho$ 
goes to zero  $K_{i\omega/\kappa}(\rho(m^2+k_j^2)^{1/2})$
behave as the delta function $\delta(\omega)$.
Thus, when one of 
the arguments of correlator (\ref{soft1}) is placed on the horizon only 
the contribution of the soft modes survives. 
By using  PV regularization and taking
the limit $\rho\rightarrow 0$ in (\ref{soft1})  one gets
$Q_{\tiny\mbox{div}}$ in form (\ref{divq}) with $R=0$.

The induced gravity proposes the following statistical-mechanical 
interpretation of black 
hole entropy formula (\ref{8c}).
$S^{BH}$ is determined by the density number
of physical states corresponding to the 
given black hole mass $M$. The physical states are related
to the states of the excitations of the constituents 
fields 
in the black hole exterior which result in the 
fluctuation of the black hole mass. The space of physical states
is obtained by factorizing 
the space of all thermal excitations over the subspace
of the soft modes. This  removes the additional
degeneracy and makes the number density of the physical states finite. 
The factorization over the soft modes
is equivalent to subtracting of the Noether
charge from the entropy $S^C$, which reduces the latter
to $S^{BH}$. 
As was pointed out in \cite{FF:97a}, there is a similarity between this 
mechanism and
gauge theories, soft modes playing the role of
the pure gauge degrees of freedom.

\section{Summary}
\setcounter{equation}0

This review is devoted to the description of thermal ensembles of
quantum fields in a spacetime of a black hole.  This study  was
stimulated by the attempts to give the statistical-mechanical
explanation  of the Bekenstein-Hawking the entropy. This is the key
problem of  black hole physics. 

The main difficulty of the statistical mechanics of quantum fields near
black holes  is connected with  additional thermal (infrared)
divergences.  In the presence of these divergences the relation between
canonical and covariant Euclidean  formulations of the theory requires
reconsideration.  In this review we payed a special attention  to the
analysis of the divergences and methods of their regularizations. It
was demonstrated  that the canonical and covariant Euclidean 
formulations are equivalent, and in the same regularization the
divergences in the both formulations are identical. An important
property of the problem is that the thermal  divergences in one
regularization take the form of ultraviolet divergences in the other
regularization. 
This duality is crucial for the discussion  of the
black hole entropy.
Thermal divergences which contribute to the black
hole entropy are connected with ultraviolet one-loop  divergences which
renormalize gravitational coupling constants.

We analyzed the problem of  black hole entropy and demonstrated it 
cannot be solved in a theory of gravity within the standard scheme of
renormalizations. The renormalization requires initial bare entropy,
which is of pure geometrical origin and (in the absence of non-minimal
coupling) is infinite negative quantity. However, if the bare classical
(tree level) gravity is absent the Bekenstein-Hawking entropy $S^{BH}$
can be directly related to the statistical-mechanical entropy $S^{C}$
of quantum black hole excitations. New important feature is that this
relation necessarily contains the Noether charge $Q$ for non-minimally
coupled fields. In one-loop ultraviolet-finite theories without bare
gravity there exist the relation between $S^{BH}$ and $S^{C}$: 
\begin{equation}\label{conc}
S^{BH}=S^{C}-Q~~~.
\end{equation}
These theories belong to the class of so-called induced gravity
theories. In these theories gravity is induced as the result of
collective quantum excitations of heavy constituents of the Planckian
mass.  The same constituent fields which generate low energy gravity
are responsible for the entropy $S^{BH}$  of a black hole. 

The analysis of concrete models shows that the relation  (\ref{conc})
is the result of the consistency of the theory. The Noether charge $Q$
determines the difference between the energy $E$ of the system and the
value of the Hamiltonian $H$. The entropy $S^{BH}$ describes the
degeneracy  of states of the black hole with respect to its mass. It
can be related to the degeneracy of the  system of constituents with
respect to its energy $E$. The latter differs from the degeneracy of
the Hamiltonian $H$. Formula (\ref{conc}) takes into account this fact.

In this approach fields (constituents) which contribute into the
entropy are to be considered as fundamental. One can expect that  such
fields  arise in the fundamental theory of quantum gravity, for
instance, in the string theory. This mechanism is not known at present,
but its possible existence might explain universality of the entropy of
black holes.

\bigskip

\vspace{12pt}
{\bf Acknowledgements}:\ \ This work was 
partially supported  by the Natural
Sciences and Engineering
Research Council of Canada. One of the authors 
(V.F.) is
grateful to the Killam Trust for its 
financial support.

\newpage
\appendix
\section{Energy, Hamiltonian 
and Noether charge}
\setcounter{equation}0

To explain the meaning of the Noether charge  $Q$ discussed in Section 7
let us
consider a classical real scalar field $\phi$ on the static 
background with the action 
\begin{equation}\label{3.8.4}
I[\phi,g]=-\frac 12\int(\phi^{,\mu}\phi_{,\mu}+m^2\phi^2+\xi
R\phi^2)\sqrt{-g}~d^4x~~~.
\end{equation}
The energy $E$ of the field in the 3D region ${\cal B}$ 
is determined 
by the stress-energy
tensor $T_{\mu\nu}$ 
\begin{equation}\label{3.8.5}
E=\int_{{\cal B}} \ T_{\mu\nu}\zeta^{\mu}d\sigma^{\nu}~~~.
\end{equation}
Here $d\sigma^{\nu}$ is the future directed vector of the volume 
element on ${\cal B}$ and $\zeta^\mu$ is the 
time-like Killing vector.
The stress-energy tensor
is obtained by the variation of action (\ref{3.8.4})
\begin{equation}\label{3.8.5a}
T_{\mu\nu}={2 \over \sqrt{-g}}{\delta I[g] \over 
\delta g^{\mu\nu}}~~~.
\end{equation}
In a static spacetime the energy $E$ is conserved
on the equations of motion of the field $\phi$ (i.e.,
$E$ does not depend on the choice of ${\cal B}$).
In addition to
the stress-energy tensor (\ref{3.8.5a}), one can 
define the {\it canonical} stress-energy tensor
\begin{equation}\label{3.8.5b}
(T^C)_{\mu\nu}=\phi_{,\mu} {\partial {\cal L} 
\over \partial \phi^{,\nu}}-g_{\mu\nu} {\cal L}~~~,
\end{equation}
where ${\cal L}$ is the Lagrangian of the theory related to
the action as $I=\int\sqrt{-g}d^4x
{\cal L}$. For static spacetimes $(T^C)_{\mu\nu}$
yields another conserved quantity which is the Hamiltonian
of the system
\begin{equation}\label{3.8.6a}
H=\int_{{\cal B}} \ T^C_{\mu\nu}\zeta^{\mu}d\sigma^{\nu}~~~.
\end{equation}
In the canonical formulation of the theory $H$ plays
the role of a generator of the evolution of the
system along the Killing time. 
In general the tensors $T_{\mu\nu}$ and $(T^C)_{\mu\nu}$
do not coincide and their difference 
yields the Noether current\footnote{The coefficient
$2\pi /\kappa$ in the definition (\ref{nc}) 
corresponds to the Killing vector $\zeta$ normalized 
so that $\zeta^2=-1$ at infinity.}
\begin{equation}\label{nc}
J_\mu={2\pi \over \kappa} \left((T^C)_{\mu\nu}-T_{\mu\nu}\right)\zeta^\nu
~~~,
\end{equation}
where $\kappa$ is the surface gravity.
According to the Noether theorem, this current 
conserves ($\nabla^\mu J_\mu=0$) on the equations
of motion. As it follows from (\ref{3.8.5})
and (\ref{3.8.6a}),
the difference between the energy $E$ and the Hamiltonian
$H$ is the Noether charge $Q$ corresponding to the
current $J_\mu$.

The simplest example when $T_{\mu\nu}$ and $(T^C)_{\mu\nu}$ are
different is the scalar field with $\xi\neq 0$.
In this case
\begin{equation}\label{3.8.7}
J_\mu=-\xi {2\pi \over \kappa}\left(R_{\mu\nu}\phi^2+g_{\mu\nu}
(\phi^2)^{,\rho}_{~;\rho}-
(\phi^2)_{;\mu\nu}\right)\zeta^\nu~~~,
\end{equation}
\begin{equation}\label{3.8.9}
H-E=\xi\int_{\partial {\cal B}}ds^k~|g_{00}|^{1/2}
\left[(\phi^2)_{,k}-\phi^2w_k\right]~~~.
\end{equation}
Here  $ds^k$  is a three dimensional vector in ${\cal B}$ normal 
to the boundary $\partial {\cal B}$ and directed outward with respect to
${\cal B}$.
Thus two energies differ by a surface term given on the
boundary $\partial {\cal B}$ of the hypersurface ${\cal B}$. Obviously,
when one considers a complete Cauchy surface 
the boundary term in (\ref{3.8.9})
contains only a contribution from the spatial infinity, or
from the external spatial boundaries if they are present.
For a field falling off 
at infinity or obeying suitable conditions at the boundary,
$E=H$. However, the situation is qualitatively
different if the integration region in $E$ is restricted by the
bifurcation surface $\Sigma$ of the Killing
horizon, where the field $\phi$ can take arbitrary finite values. If the
contribution from the spatial infinity or external boundary is absent
only the Nether charge $Q$ on $\Sigma$ gives the contribution
to the difference $H-E$
\begin{equation}\label{3.8.10x}
H-E={\kappa \over 2\pi}Q~~~,
\end{equation}
\begin{equation}\label{2.10a}
Q=2\pi \xi \int_{\Sigma}\sqrt{\sigma}
d^2\theta~\phi^2~~~.
\end{equation}
In an analogous way one can compute the charge $Q$
in other theories. For spin 1/2 fields a trivial
check shows that the energy and Hamiltonian are
identical and $Q=0$.
Presumably, $Q$ is non-trivial in the theories 
which include fields with
higher spins.


\begin{thebibliography}{000}

\bibitem{BCH:73} J.M. Bardeen, B. Carter, and 
S.W. Hawking, Commun. Math. Phys. {\bf 31}
(1973) 161.

\bibitem{MTW:73} C.W. Misner, K.S. Thorne, and
J.A. Wheeler, {\it Gravitation}, San Francisco:
Freeman, 1973.

\bibitem{Bekenstein:72} J.D. Bekenstein, Nuov. Cim. Lett. {\bf 4}   
(1972) 737. 

\bibitem{Bekenstein:73} J.D. Bekenstein,
Phys. Rev. {\bf D7} (1973) 2333.  

\bibitem{Bekenstein:74} J.D. Bekenstein,
Phys. Rev. {\bf D9}  (1974) 3292.

\bibitem{Hawking:75} S.W. Hawking, Comm. Math. Phys.
{\bf 43} (1975) 199.

\bibitem{StVa:96} A. Strominger and C. Vafa, 
Phys. Lett. {\bf B379} (1996) 99.

\bibitem{JKM:96} C.V. Johnson, R.R. Khuri, and R.C. Myers,
Phys. Lett. {\bf 378} (1996) 78.

\bibitem{MaSt:96} J.M. Maldacena and A. Strominger,
Phys. Rev. Lett. {\bf 77} (1996) 428.

\bibitem{CaMa:96} C.G. Callan and J.M. Maldacena,
Nucl. Phys. {\bf B472} (1996) 591.

\bibitem{HoSt:96} G.T. Horowitz and A. Strominger,
Phys. Rev. Lett. {\bf 77} (1996) 2368.

\bibitem{Horo:97} G.T. Horowitz, {\it Quantum States
of Black Holes}, preprint gr-qc/9704072.

\bibitem{Peet:97} A.W. Peet, {\it The Bekenstein Formula and String 
Theory (N-Brane Theory)}, preprint hep-th/9712253.

\bibitem{Carlip:95} S. Carlip, Phys. Rev. {\bf D51} (1995) 632.

\bibitem{Strominger:97} A. Strominger, {\it Black Hole Entropy from Near 
Horizon Microstates}, preprint hep-th/9712251.

\bibitem{ABCK:97} A. Ashtekar, J. Baez, A. Corichi, K. Krasnov,
{\it Quantum Geometry and Black Hole Entropy}, preprint gr-qc/9710007.

\bibitem{ZuTh:85} W.H. Zurek and K.S. Thorne,
Phys. Rev. Lett. {\bf 54} (1985) 2171.

\bibitem{Hooft:85} G.'t Hooft, Nucl. Phys. {\bf B256} (1985) 727.

\bibitem{TPM:86} K.S. Thorne, R.H. Price, and D.A. Macdonald,
{\it Black Holes: The Membrane Paradigm}, Yale University Press
1986, New Haven and London.


\bibitem{Israel:76} W. Israel, Phys. Lett.
{\bf 57A} (1976) 107.

\bibitem{BiWi:76} J.J. Bisognano and E.H. Wichmann, 
J. Math. Phys. {\bf 17} (1976) 303.

\bibitem{UnWe:84} W.G. Unruh and N. Weiss,
Phys. Rev. {\bf D29} (1984) 1656.

\bibitem{Takagi:86} S. Takagi, Progress of Theoretical
Physics Supplement {\bf 88} (1986).

\bibitem{KaWa:91} B.S. Kay and R.M. Wald, 
Phys. Rep. {\bf 207}(2) (1991) 49.

\bibitem{Laflamme:89} R. Laflamme, Nucl. Phys. {\bf B324}
(1989) 233.


\bibitem{BKLS:86} L. Bombelli, R.K. Koul, J. Lee, 
and R.Sorkin, Phys. Rev. {\bf D34}  (1986) 373.

\bibitem{Srednicki:93} M. Srednicki, Phys.  Rev. Lett.
{\bf 71} (1993) 666.

\bibitem{MuSeKo:97} S. Mukohyama, M. Seriu and H. Kodama,
Phys. Rev. {\bf D55} (1997) 7666. 


\bibitem{LaWi:95} F. Larsen and F. Wilczek, 
Ann. Phys. {\bf 243} (1995) 280.

\bibitem{BePi:96} E. Benedict and S.-Y. Pi,
Ann. Phys. {\bf 245} (1996) 209.


\bibitem{FrNo:93} V. Frolov and I. Novikov, Phys. Rev. 
{\bf D48} (1993) 4545.

\bibitem{Bekenstein:94} J.D. Bekenstein, 
{\it Do We Understand Black Hole Entropy?}, 
7th Marcel Grossman Meeting on General Relativity
(1994) p. 39, gr-qc/9409015.


\bibitem{KaSt:94} D. Kabat and M.J. Strassler,
Phys. Lett. {\bf B329} (1994) 46.

\bibitem{CaWi:94} C. Callan and F. Wilczek, 
Phys. Lett {\bf B333} (1994) 55.

\bibitem{Jacobson:94b} T. Jacobson, Phys. Rev. {\bf D50}
(1994) 6031.

\bibitem{BFZ:95} A.O. Barvinsky, V.P. Frolov, 
and A.I. Zelnikov, Phys. Rev. {\bf D51} (1995) 1741.

\bibitem{Frolov:95} V.P. Frolov, Phys. Rev. Lett.
{\bf 74} (1995) 3319.

\bibitem{SuUg:94} L. Susskind and J. Uglum, Phys. Rev.
{\bf D50} (1994) 2700.

\bibitem{DLM:95} J.-G. Demers, R. Lafrance, and R.C. Myers,
Phys. Rev. {\bf D52}  (1995) 2245.

\bibitem{FS:96} D.V. Fursaev and S.N. Solodukhin,
Phys. Lett. {\bf B365} (1996) 51.

\bibitem{Solodukhin:95} S.N. Solodukhin, Phys. Rev.
{\bf D52} (1995) 7046.

\bibitem{LaWi:96} F. Larsen and F. Wilczek,
Nucl. Phys. {\bf B458} (1996) 249.

\bibitem{KSS:95} D. Kabat, S.H. Shenker, and 
M.J. Strassler, Phys. Rev. {\bf D52} (1995) 7027.

\bibitem{HKN:97} M. Hotta, T. Kato, and K. Nagata,
Class. Quantum Grav. {\bf 14} (1997) 1917.


\bibitem{Jacobson:94} T. Jacobson, {\it Black Hole
Entropy in Induced Gravity}, gr-qc/9404039.

\bibitem{Sakh:68} A.D. Sakharov, Sov. Phys.
Doklady {\bf 12} (1968) 1040.

\bibitem{Sakh:76} A.D. Sakharov, Theor. Math. Phys.
{\bf 23} (1976) 435.

\bibitem{FFZ:97} V.P. Frolov, D.V. Fursaev, and A.I. Zelnikov, Nucl. 
Phys. {\bf B486} (1997) 339.

\bibitem{FF:97a} V.P. Frolov and D.V. Fursaev,
Phys. Rev. {\bf D56} (1997) 2212.

\bibitem{FF:97b} V.P. Frolov and D.V. Fursaev,
{\it Plenty of Nothing: Black Hole Entropy
in Induced Gravity}, hep-th/9703178.



\bibitem{Gibbons:77} G.W. Gibbons, Phys. Lett.
{\bf 60A} (1977) 385.

\bibitem{Gibbons:78} G.W. Gibbons, in {\it Differential
Geometrical Methods in Mathematical Physics II}
edited by K. Bleuler, H.R. Petry, and A. Reetz
(Springer, New York, 1978), p. 518.

\bibitem{GiPe:78} G.W. Gibbons and M.J. Perry,
Proc. R. Soc. Lond. {\bf A358} (1978) 467.

\bibitem{DoKe:78} J.S. Dowker and G. Kennedy, 
J. Phys. A: Math. Gen. {\bf 11} (1978) 895.

\bibitem{York:86} J.W.~York,  Phys. Rev. 
{\bf D33} (1986) 2092.

\bibitem{LaLi} L. Landau and I. Lifshitz,
{\it Statistical physics}, v. 1, Oxford;
New York: Pergamon Press. 1980.


\bibitem{Allen:86} B. Allen, Phys. Rev. {\bf D33} 
(1986) 3640.

\bibitem{GiHa:76} G.W.~Gibbons and S.W.~Hawking,  Phys. Rev. {\bf D15} 
(1976) 2752.

\bibitem{Hawk:79} S.W.~Hawking, In: {\em General Relativity: An
Einstein Centenary Survey.} (eds. S.W.~Hawking and W.~Israel),
Cambridge Univ.Press, Cambridge, 1979.



\bibitem{BCMMWY:90} J.D. Brown, G.L. Comer, E.A. Martinetz, J. Malmed,
B.F. Whiting and J.W. York, Class. Quantum Grav.  {\bf 7} (1990) 1433.

\bibitem{BBWY:90} H.~W.~Braden, J.~D.~Brown, B.~F.~Whiting, and
J.~W.~Jork, Phys. Rev. {\bf D42} (1990) 3376.


\bibitem{Dowker:84} J.S. Dowker, Class. Quantum. Grav. {\bf 1} (1984)
369.

\bibitem{Dowker:89} J.S. Dowker, Phys. Rev. {\bf D39}
(1989) 1235.

\bibitem{DoSc:88} J.S. Dowker and J.P. Schofield,
Phys. Rev. {\bf D38} (1988) 3327.

\bibitem{DoSc:89} J.S. Dowker and J.P. Schofield,
Nucl. Phys. {\bf 327} (1989) 267.


\bibitem{Fursaev:97} D.V. Fursaev, {\it 
Euclidean and Canonical Formulations of 
Statistical Mechanics in 
the Presence of Killing Horizons}, hep-th/9709213.



\bibitem{BiDa:82} N.D. Birrell and P.C.W. Davies, {\it Quantum Fields  
in Curved Space}, Cambridge University Press, Cambridge
1982.

\bibitem{Duff:77} M.J. Duff, Nucl. Phys. {\bf 125}
(1977) 334.

\bibitem{Fujikawa:80} K. Fujikawa, Phys. Rev. Lett.
{\bf 44} (1980) 1733.

\bibitem{Fujikawa:81} K. Fujikawa, Phys. Rev. 
{\bf D23} (1981) 2262.

\bibitem{FrVi:73} E.S. Fradkin and G.A. Vilkoviskii,
Phys. Rev. {\bf D8} (1973) 4241.

\bibitem{Hawking:77} S.W. Hawking, Commun. Math. Phys. 
{\bf 55} (1977) 133.


\bibitem{deAOh:95} S.P. de Alwis and N. Ohta,
{\bf D52} (1995) 3529.


\bibitem{CVZ:95a} G. Cognola, L. Vanzo and S. Zerbini,
Class. Quantum Grav. {\bf 12} (1995) 1927.

\bibitem{BCZ:96} A.A. Bytsenko, G. Cognola, and
S. Zerbini, Nucl. Phys. {\bf B458} (1996) 267.


\bibitem{HaHa:76} J.B. Hartle and S.W. Hawking,
Phys. Rev. {\bf D13} (1976) 2188.


\bibitem{Bateman:54} H. Bateman and A. Erdelyi,
{\it Tables of Integral Transformations}, v.1,
New York, McGraw-Hill Book Company, Inc., 1954.


\bibitem{Camporesi:90} R. Camporesi, Phys. Rep. 
{\bf 196} (1990) 1.

\bibitem{BCVZ:96} A.A. Bytsenko, G. Cognola, 
L. Vanzo, and S. Zerbini, Phys. Rep. {\bf 266}
(1996) 1.

\bibitem{FFZ:96a} V.P. Frolov, D.V. Fursaev 
and A.I. Zelnikov, Phys. Rev. {\bf D54} (1996) 2711.

\bibitem{ZCV:96} S. Zerbini, G. Cognola, and
L. Vanzo, Phys. Rev. {\bf D54} (1996) 2699.


\bibitem{MTZ:92} R.B. Mann, L. Tarasov, and A. Zelnikov,
Class. Quantum Grav. {\bf 9} (1992) 1487.


\bibitem{Solod:96a} S.N. Solodukhin,
Phys. Rev. {\bf D54} (1996) 3900.

\bibitem{Romeo:96} A. Romeo, Class. Quantum Grav.
{\bf 13} (1996) 2797.

\bibitem{KKSY:97} S.P. Kim, S.K. Kim, K.-W. Soh,
J.H. Yee, Phys. Rev. {\bf D55} (1997) 2159.


\bibitem{Solod:96b} S.N. Solodukhin, Phys. Rev. {\bf D56} 
(1997) 4968.

\bibitem{CoLe:97} G. Cognola and P. Lecca, Phys. Rev.
{\bf D57} (1998) 1108.


\bibitem{SCZ:97} Y.-G. Shen, D.-M. Chen and T.-J. Zhang,
Phys. Rev. {\bf D56} (1997) 6698.


\bibitem{NaFu:85} N. Nakazawa and T. Fukuyama,
Nucl. Phys. {\bf B252} (1985) 621.

\bibitem{GrRy:94} I.S. Gradshteyn and I.M. Ryzhik,
{\it Table of Integrals, Series, and Products},
Academic Press, New York 1994.

\bibitem{GuZe:97} Yu. Gusev and A. Zelnikov, 
{\it Nonlocal Effective Action at Finite Temperature in
Ultrastatic Space-Times}, preprint hep-th/970974.


\bibitem{FSW:96} V.P. Frolov, W. Israel, and
S.N. Solodukhin, Phys. Rev. {\bf D54} (1996) 2732.

\bibitem{FFZ:96b} V.P. Frolov, D.V. Fursaev 
and A.I. Zelnikov, Phys. Lett. {\bf B382} (1996) 220.

\bibitem{MaSo:97} R.B. Mann and S.N. Solodukhin,
Phys. Rev. {\bf D55} (1997) 3622.

\bibitem{FM:94} D.V. Fursaev and G. Miele, Phys. Rev. 
{\bf 49} (1994)  
987.

\bibitem{DFM:97b}  L. De Nardo, D.V. Fursaev and G. Miele,
Class. Quantum Grav. {\bf 14} (1997) 3269.

\bibitem{Vilenkin:85} A.A. Vilenkin, 
Phys. Rep. {\bf 121} N5, 263 (1985).


\bibitem{GSW:87} M.B. Green, J.H. Schwarz and E. Witten, {\it Superstring
Theory}, vol. 2, Cambridge University Press, Cambridge, 1987.

\bibitem{Volovik:97} G.E. Volovik, {\it Simulation of
Quantum Field Theory and Gravity in Superfluid He-3},
preprint cond-mat/9706172.


\bibitem{Cheeger:83} J. Cheeger, J. Differential
Geometry, {\bf 18} (1983) 575.

\bibitem{Donnelly:76} H. Donnelly, Math. Ann. {\bf 224}
(1976) 161.

\bibitem{KaSt:91} B.S. Kay and U.M. Studer,
Commun. Math. Phys. {\bf 139} (1991) 103.

\bibitem{Dowker:77} J.S. Dowker, J. Phys. A: Math. Gen. {\bf 10} (1977) 115.

\bibitem{Dowker:87} J.S. Dowker, Phys. Rev. {\bf D15}
(1987) 3742.

\bibitem{DeJa:88} S. Deser and R. Jackiw, 
Comm. Math. Phys {\bf 118} (1988) 495.

\bibitem{Sommerfeld:1897} A. Sommerfeld, Proc.
Lond. Math. Soc. {\bf 28} (1897) 417.


\bibitem{CKV:94} G. Cognola, K. Kirsten and L. Vanzo, Phys. Rev. {\bf  
D49} (1994) 1029.

\bibitem{Fursaev:94a} D.V. Fursaev, Class. Quantum
Grav. {\bf 11} (1994) 1431.

\bibitem{Kabat:95} D. Kabat, Nucl. Phys. {\bf B453} (1995) 281.

\bibitem{FM:97} D.V. Fursaev and G. Miele, 
Nucl. Phys. {\bf B484} (1997) 697.



\bibitem{Fursaev:94b} D.V. Fursaev, Phys. Lett. {\bf B334} (1994) 53.

\bibitem{Dowker:94a} J.S. Dowker,
Class. Quantum Grav. {\bf 11} (1997) L137.

\bibitem{Dowker:94b} J.S. Dowker, Phys. Rev. {\bf D50}
(1994) 6369.

\bibitem{DFM:97a} L. De Nardo, D.V. Fursaev and G. Miele,
Class. Quantum Grav. {\bf 14} (1997) 1059.


\bibitem{MoIe:97} V. Moretti and D. Iellici,
Phys. Rev. {\bf D55} (1997) 3552.

\bibitem{MoIe:96} V. Moretti and D. Iellici,
Phys. Rev. {\bf D54} (1996) 7459.

\bibitem{Moretti:97} V. Moretti, J. Math. Phys.
{\bf 38} (1997) 2922.

\bibitem{MaSo:96} R.B. Mann and S.N. Solodukhin,
Phys. Rev. {\bf D54} (1996) 3932.

\bibitem{IcSa:95} I. Ichinose and Yu. Satoh, Nucl. Phys. {\bf B447}
(1995) 340.

\bibitem{Satoh:97} Yu. Satoh, {\it Study of Three-dimensional Quantum 
Black Holes}, hep-th/9705209.

\bibitem{LeKi:96} M.-H. Lee and J.K. Kim, Phys. Rev. {\bf D54} (1996)
(3904).

\bibitem{LKK:96} M.-H. Lee, H.-C. Kim, and J.K. Kim, Phys. Lett. {\bf
B388} (1996) 487.

\bibitem{HKP:97} J. Ho, W.T. Kim, and Y.-J. Park,
Class. Quantum Grav. {\bf 14} (1997) 2617.

\bibitem{AHL:95} P.R. Anderson, W.A. Hiscock,
and D.J. Loranz, Phys. Rev. Lett. {\bf 74} (1995) 4365.

\bibitem{Vanzo:97} L. Vanzo, Phys. Rev. {\bf D55} 
(1997) 2192.

\bibitem{HHR:95} S.W. Hawking, G. Horowitz, and
S. Ross, Phys. Rev. {\bf D51} (1995) 4302.

\bibitem{Teitelboim:95} C. Teitelboim, Phys. Rev.
{\bf D51} 4315.

\bibitem{GhMi:95} A. Ghosh and P. Mitra, Phys. Lett.
{\bf B357} (1995) 295.

\bibitem{CVZ:95b} G. Cognola, L. Vanzo and S. Zerbini,
Phys. Rev. {\bf D52} (1995) 4548.


\bibitem{FS:95} D.V. Fursaev and S.N. Solodukhin,
Phys. Rev. {\bf D52} (1995) 2133.

\bibitem{Wald:93} R.M. Wald, Phys. Rev. {\bf D48}
(1993) R3427.

\bibitem{IyWa:94} V. Iyer and R.M. Wald,
Phys. Rev. {\bf D50} (1994) 846.

\bibitem{JKM:94} T.A. Jacobson, G. Kang, and R.C. Myers,
Phys. Rev. {\bf D49} (1994) 6587.

\bibitem{Emparan:95} R. Emparan, Phys. Rev. 
{\bf D51} (1995) 5716.

\bibitem{BaEm:95} J.L.F. Barbon and R. Emparan,
Phys. Rev. {\bf 52} (1995) 4527.

\bibitem{BeLi:96} F. Belgiorno and S. Liberati,
Phys. Rev. {\bf D53} (1996) 3172.

\bibitem{BeMa:96} F. Belgiorno and M. Martellini,
Phys. Rev. {\bf D53} (1996) 7073.

\bibitem{Solod:95b} S.N. Solodukhin, Phys. Rev. {\bf D51}
(1995) 609, ibid 618.

\bibitem{Fursaev:95} D.V. Fursaev, Mod. Phys. Lett. 
{\bf A10} (1995) 649.

\bibitem{Nelson:94} W. Nelson, Phys. Rev. {\bf D50}
(1994) 7400.

\bibitem{Brown:95} J.D. Brown, Phys. Rev. {\bf D52}
(1995) 7011.


\bibitem{Adler:82} S.L. Adler, Rev. Modern Phys.
{\bf 54} (1982) 729.

\bibitem{NoVa:91} Yu.V. Novozhilov and D.V. Vassilevich,
Lett. Math. Phys. {\bf 21} (1991) 253.


\bibitem{MoIe:97b} V. Moretti and D. Iellici,
{\it $\zeta$-function regularization and one-loop 
renormalization of field fluctuations in curved
spacetimes}, preprint-UTF 401, hep-th/9705077.

\bibitem{FN:book} I.D. Novikov and V.P. Frolov,
{\it Physics of Black Holes}, Kluwer Academic 
Publishers, Dordrecht, 1989.



\end{thebibliography}
\end{document}